\documentstyle[eqsecnum,aps]{revtex}

\begin{document}
\draft
\title{Evolution of the scale factor with a variable
       cosmological term}
\author{J. M. Overduin\cite{byline} and F. I. Cooperstock}
\address{Department of Physics and Astronomy\\
         University of Victoria\\
         Victoria, British Columbia, Canada V8W 3P6}
\date{\today}
\maketitle
\begin{abstract}
Evolution of the scale factor $a(t)$ in Friedmann models
(those with zero pressure and a constant cosmological term $\Lambda$)
is well understood, and elegantly summarized in the review of 
Felten and Isaacman [Rev. Mod. Phys. 58, 689 (1986)].
Developments in particle physics and inflationary theory,
however, increasingly indicate that $\Lambda$ ought to be
treated as a dynamical quantity.  
We revisit the evolution of the scale factor with a variable $\Lambda$-term,
and also generalize the treatment to include nonzero pressure.
New solutions are obtained and evaluated using a variety of
observational criteria.
Existing arguments for the inevitability of a big bang 
(ie., an initial state with $a=0$) are substantially weakened,
and can be evaded in some cases with $\Lambda_0$ 
(the present value of $\Lambda$) 
well below current experimental limits.
\end{abstract}


\section{Introduction}
\label{sec:int}

The behavior of the cosmological scale factor $a(t)$ in solutions
of Einstein's field equations with the Robertson-Walker line
element has been the subject of numerous studies.
Textbook presentations tend to focus on models in which
pressure $p$ is zero and there is no cosmological term ($\Lambda=0$).
Some treatments include the
{\em Friedmann models\/}, in which $p=0$ but $\Lambda \neq 0$
\cite{rob33,ref67,edw72,rin77,fel86}.
Much less attention has been directed at the more general
{\em Lema\^{\i}tre models\/}, in which pressure $p$ is given in terms
of density $\rho$ by an equation of state $p=p(\rho)$,
and $\Lambda \neq 0$ \cite{mcv65,har67,mci68,agn70}.

The possibility of a nonzero $\Lambda$-term, in particular,
has resurfaced lately in connection with the age problem \cite{Age}.
If $\Lambda$ is large enough, in fact, the age of the Universe
(defined in the standard model \cite{wei72} as the time elapsed
since $a=0$) can in principle become {\em infinite\/}, 
as $a(t)$ never drops below a nonzero minimum value $a_{\ast}$ 
in the past direction.  
The existence of such ``big bangless'' oscillating models has 
been recognized for over sixty years \cite{Discuss}.
They have however been dismissed as unphysical on the grounds that
the required values of $\Lambda$ are incompatible with observation 
\cite{ell84,bor88,lah91,Note}.

We will return to the observational constraints later on;
merely noting here that the above arguments, along with nearly all
existing astrophysical work on the cosmological term,
operate on the assumption that $\Lambda=$~{\em constant\/}.
Quantum field theorists and others, by contrast, have been treating 
the cosmological term as a {\em dynamical\/} quantity for thirty
years \cite{ber68,lin74,end77,can77,kaz80,pol82}.
Anything which contributes to the energy density
$\rho_v$ of the vacuum behaves like a cosmological term
via $\Lambda_v=8\pi G \rho_v$.
Many potential sources of fluctuating vacuum energy have now been
identified, including scalar fields \cite{dol83,Scalar},
tensor fields \cite{Tensor,dol97}, nonlocal effects \cite{Nonlocal},
wormholes \cite{Worm}, inflationary mechanisms \cite{Inflate}
and even cosmological perturbations \cite{Pert}.
Each of these has been claimed to give rise to a negative energy 
density which grows with time, tending to cancel out any pre-existing positive 
cosmological term and drive the net value of $\Lambda$ toward zero.
Processes of this kind are among the most promising ways to resolve the
longstanding cosmological ``constant'' problem \cite{dol97}
(see \cite{wei89} for review).

The purpose of the present paper is to re-examine the 
evolution of the scale factor when $\Lambda\neq$~constant.
This has not yet been done in a systematic way.
We also expand on most earlier treatments by considering
a fairly general equation of state for ordinary matter rather than 
restricting ourselves to the pressure-free Friedmann models.
These two generalizations lead to qualitatively as well as
quantitatively new behavior for $a(t)$, and hence for
related phenomena such as the age of the Universe,
the appearance of gravitational lenses,
and the redshifts of distant astronomical objects.
They can also allow one to circumvent the abovementioned 
arguments against oscillating models.

The remainder of the paper is organized as follows: 
the required assumptions, definitions and dynamical equations are 
assembled in \S\ref{sec:prelim}, 
and applied in \S\S\ref{sec:t} -- \ref{sec:q} to models in which 
$\Lambda$ varies as a function of time $t$, the scale factor $a$,
the Hubble parameter $H$ and the deceleration parameter $q$
respectively.  
In each section we obtain analytic or numerical solutions for the
scale factor, paying particular attention to the question of
the initial singularity, and discuss observational constraints
where appropriate.  Conclusions are summarized in \S\ref{sec:con}.

\section{Variable-{\boldmath $\Lambda$} Cosmology}
\label{sec:prelim}

\subsection{The Cosmological ``Constant''}

To begin, we recall why the cosmological term has often been 
treated as a constant of nature.  The Einstein field equations read:
\begin{equation}
G_{\mu\nu} + \Lambda \, g_{\mu\nu} = 8\pi G T_{\mu\nu} ,
\label{EFEs}
\end{equation}
where $G_{\mu\nu} \equiv R_{\mu\nu}-R g_{\mu\nu}/2$ is the Einstein tensor
and $T_{\mu\nu}$ is the energy-momentum tensor of matter 
(we assume units such that $c=1$).
Taking the covariant divergence of eq.~(\ref{EFEs}), recalling that
the vanishing covariant divergence of the Einstein tensor is guaranteed
by the Bianchi identities, and  assuming that the energy-momentum tensor 
satisfies the conservation law $\nabla^{\nu}T_{\mu\nu}=0$, it follows
that the covariant divergence of $\Lambda g_{\mu\nu}$ must vanish also,
and hence that $\Lambda=$~constant.
This argument, which situates $\Lambda$ firmly 
on the left-hand side of the field equations, constitutes a 
``geometrical interpretation'' of the cosmological term.

More recently, it has become increasingly common (see, eg., \cite{mis73})
to move the cosmological term to the right-hand side of eq.~(\ref{EFEs}):
\begin{equation}
G_{\mu\nu} = 8\pi G \tilde{T}_{\mu\nu} ,
\end{equation}
where:
\begin{equation}
\tilde{T}_{\mu\nu} \equiv T_{\mu\nu} - \frac{\Lambda}{8\pi G} \, g_{\mu\nu} ;
\label{effEMT}
\end{equation}
that is, to interpret $\Lambda$ as part of the matter content of the 
universe, rather than a purely geometrical entity.
Once this is done, there are no {\em a priori\/} reasons why $\Lambda$
should not vary --- as long as the {\em effective\/} 
energy-momentum tensor $\tilde{T}_{\mu\nu}$ satisfies energy 
conservation \cite{Aside}.

\subsection{Dynamical Equations}
\label{sec:dyn}

We will make the usual assumptions:  a homogeneous and isotropic
Universe (ie., Robertson-Walker line element) and perfect-fluid-like
ordinary matter \cite{Absorb} with pressure $p$ and energy density $\rho$.
Eq.~(\ref{effEMT}) then implies that the {\em effective\/} 
energy-momentum tensor also has the perfect fluid form, 
with effective pressure
$\tilde{p} \equiv p-\Lambda/8\pi G$ and energy density
$\tilde{\rho} \equiv \rho+\Lambda/8\pi G$ \cite{wei72}.
The field equations~(\ref{EFEs}) and law of energy conservation then read:
\begin{eqnarray}
\dot{a}^2 = \frac{8\pi G}{3} \rho a^2 & + & \frac{\Lambda}{3} a^2 - k
\label{LemEq} \\
\frac{d}{da} \left[ \left( \rho + \frac{\Lambda}{8\pi G} \right) a^3 \right] 
             = & - & 3 \left( p - \frac{\Lambda}{8\pi G} \right) a^2 .
\label{ECE}
\end{eqnarray}
For the equation of state we take:
\begin{equation}
p = ( \gamma - 1 ) \rho ,
\label{eqState}
\end{equation}
with $\gamma=$~constant.
Previous analytic and numerical studies of the evolution of the scale factor 
have tended to focus on the dust-like case $\gamma=1$ 
\cite{ref67,edw72,rin77,fel86}, and occasionally also the
radiation-like case $\gamma=4/3$ \cite{mcv65,har67,mci68,agn70,lan75}.
We will not restrict ourselves to these values, as many other 
possibilities have been considered in the literature \cite{Gamma}.

Substituting eq.~(\ref{eqState}) into eq.~(\ref{ECE}), we find:
\begin{equation}
\frac{d}{da} \left( \rho a^{3\gamma} \right) = -\frac{a^{3\gamma}}{8\pi G}
                                                \frac{d\Lambda}{da} .
\label{genECE}
\end{equation}
When $\Lambda=$~constant, eq.~(\ref{genECE}) reverts to the well-known result
that density $\rho$ scales as $a^{-3}$ in a pressure-free universe ($\gamma=1$)
and $a^{-4}$ in a radiation-dominated universe ($\gamma=4/3$).

Differentiating eq.~(\ref{LemEq}) with respect to time and inserting
eq.~(\ref{genECE}), we obtain:
\begin{equation}
\ddot{a} = \frac{8\pi G}{3} \left( 1 - \frac{3\gamma}{2} \right) \rho a +
           \frac{\Lambda}{3} \, a .
\label{accel}
\end{equation}
Eq.~(\ref{accel}) shows that a positive density $\rho$ acts to decelerate 
the expansion, as expected --- but only when $\gamma > 2/3$.
If, on the other hand, the cosmological fluid is such that $\gamma < 2/3$, 
then its density can actually accelerate the expansion, $\ddot{a}>0$
(this is the phenomenon commonly known as {\em inflation\/}).
Eq.~(\ref{accel}) also confirms that a positive cosmological term
contributes positively to acceleration, ``propping up''
the scale factor against the deceleration caused by the matter term $\rho$.
A negative $\Lambda$-term, on the other hand, acts in
the opposite direction and brings about recollapse more quickly.

Combining eqs.~(\ref{LemEq}) and (\ref{accel}), we can eliminate $\rho$ 
to obtain a differential equation for the scale factor in terms of the 
cosmological term alone:
\begin{equation}
\frac{\ddot{a}}{a} = \left( 1 - \frac{3\gamma}{2} \right)
                     \left( \frac{\dot{a}^2}{a^2} + \frac{k}{a^2} \right)
                   + \frac{\gamma}{2} \Lambda .
\label{mainDE}
\end{equation}
This differential equation governs the behavior of the scale factor 
in the presence of a cosmological term $\Lambda$, whether or not the
latter is constant.

\subsection{Phenomenological {\boldmath $\Lambda$}-Decay Laws}
\label{sec:phenom}

The abovementioned sources of negative vacuum energy
\cite{dol83,Scalar,Tensor,dol97,Nonlocal,Worm,Inflate,Pert}
do not, in general, lend themselves to simple expressions for
$\Lambda$ in terms of $t,a,H$ or $q$.  
There are some exceptions, including scalar field-based
\cite{end77,can77,ber86,wet95} and other theories
\cite{raj83,mai94,mof96,hoy97,joh97} in which analytic decay laws 
are derived from modified versions of the Einstein action.
In most such papers, however, no exact solution for $\Lambda$ 
is obtained; the intent is merely to demonstrate that decay 
(and preferably near-cancellation) of the effective cosmological term 
is possible in principle.

In a complementary approach, a number of authors have constructed
models of a more phenomenological character, in which specific decay
laws are postulated for $\Lambda$ within general relativity.
These theories are incomplete to the extent that they do not include
explicit physical mechanisms to govern the exchange of energy between
the shrinking cosmological term and other forms of matter \cite{rat88}.
In some models this issue is not addressed at all;
in others the energy is assumed to be channelled into
production of baryonic matter and/or radiation.
The former case can be constrained by observations of the 
diffuse gamma-ray background, since (assuming the decay process does 
not violate baryon number) one might expect equal amounts of matter and 
antimatter to be formed \cite{fre87,mat95}.  The latter can be 
constrained by nucleosynthesis arguments \cite{fre87,sat90},
cosmic microwave background (CMB) anisotropies \cite{fre87,sil94,sil97},
the absolute CMB intensity \cite{ove93,mat95}, and
thermodynamical considerations \cite{pav91,sal93,lim96}.

We take the point of view here that simple $\Lambda$-decay scenarios
are worth examining, irrespective of whether they come from extended 
theories of gravity or phenomenological considerations,
for at least four reasons:
(1) they have been shown to address a number of pressing problems
in cosmology ($\! \!$ \cite{lau85} -- \cite{rat88});
(2) many are independently motivated, eg., by dimensional arguments,
or as limiting cases of more complicated theories;
(3) most are simple enough that meaningful conclusions can be drawn
about their viability; and
(4) successful implementation would point toward 
the eventual Lagrangian formulation of a more complete theory.
For convenience, we have collected together the most common decay
laws from the literature and listed them in Table~\ref{table1}
(by chronological order of appearance).

In the remainder of the paper, we focus on power-law functions of one 
parameter.  Our discussion is however more general than most of those 
noted in Table~\ref{table1}, because we do not fix values of the exponents
{\em a priori\/}.  In particular, we consider decay laws of the 
following four kinds:
\begin{eqnarray}
\Lambda & = & {\cal A} \, t^{-\ell} \label{t-Law} \\
\Lambda & = & {\cal B} \, a^{-m} \label{a-Law} \\
\Lambda & = & {\cal C} \, H^{\, n} \label{H-Law} \\
\Lambda & = & {\cal D} \, q^{\, r} \label{q-Law} ,
\end{eqnarray}
where ${\cal A,B,C,D},\ell,m,n$ and $r$ serve as adjustable constants.
These four variable-$\Lambda$ scenarios and their cosmological
consequences are explored in \S\ref{sec:t}, \S\ref{sec:a}, 
\S\ref{sec:H} and \S\ref{sec:q} respectively.

\subsection{Definitions}
\label{sec:defns}

We conclude \S\ref{sec:prelim} by introducing the terms and definitions which
will be needed to connect our solutions to observation.  Chief among these
are the energy densities of ordinary matter and the cosmological term,
expressed in units of the critical density:
\begin{equation}
     \Omega \equiv \frac{\rho}{\rho_{crit}} \; \; \; , \; \; \;
    \lambda \equiv \frac{\Lambda}{3H_0^2} \; \; \; , \; \; \;
\rho_{crit} \equiv \frac{3H_0^2}{8\pi G} .
\label{lamOmDefns}
\end{equation}
We will be particularly interested in the values of these parameters
at the present time (subscript ``0''):
\begin{equation}
 \Omega_0 = \frac{8\pi G \rho_0}{3H_0^2} \; \; \; , \; \; \;
\lambda_0 = \frac{\Lambda_0}{3H_0^2} .
\label{lamOm0defns}
\end{equation}
These will constitute our primary free parameters throughout the 
following sections.

The usefulness of the quantity $\lambda_0$ 
(sometimes denoted $\Omega_{\Lambda}$ in the literature)
is highlighted by
using the definitions~(\ref{lamOmDefns}) to rewrite the Lema\^{\i}tre 
equation~(\ref{LemEq}) in the form $H^2=H_0^2(\Omega +\lambda)-k/a^2$.
At the present time this implies:
\begin{equation}
k = a_0^2 H_0^2 ( \Omega_0 + \lambda_0 - 1 ) .
\label{Useful}
\end{equation}
From eq.~(\ref{Useful}) it is clear that:
\begin{eqnarray}
\Omega_0 + \lambda_0 > 1 \hspace{5mm} & \Longrightarrow & \hspace{5mm}
                   k > 0 \nonumber \\
\Omega_0 + \lambda_0 = 1 \hspace{5mm} & \Longrightarrow & \hspace{5mm}
                   k = 0 \nonumber \\
\Omega_0 + \lambda_0 < 1 \hspace{5mm} & \Longrightarrow & \hspace{5mm}
                   k < 0 .
\label{kDefn}
\end{eqnarray}
Most cosmologists implicitly choose units for $a_0$ such that
the value of $k$ is normalized to either 0 or $\pm1$.  We will follow
Felten and Isaacman \cite{fel86} in refraining from this, because it is
more convenient for our purposes to choose units such that $a_0 \equiv 1$.

In place of $\Omega$ and $\lambda$, some authors prefer to use the
quantities $\sigma$ and $q$, defined by:
\begin{equation}
\sigma \equiv \frac{4\pi G \rho}{3H^2} \; \; \; , \; \; \;
     q \equiv -\frac{\ddot{a}a}{\dot{a}^2} .
\label{qDefn}
\end{equation}
The present values of these parameters are fixed with the help of
eqs.~(\ref{accel}) and(\ref{lamOmDefns}):
\begin{equation}
\sigma_0 = \Omega_0/2 \; \; \; , \; \; \;
     q_0 = (3\gamma/2 - 1) \Omega_0 - \lambda_0 .
\label{q0defn}
\end{equation}
Negative values of $q_0$ can be obtained for large (positive) $\lambda_0$,
or small $\gamma$.  Eq.~(\ref{q0defn}) yields the standard
expression \cite{fel86} for $q_0$ when $\gamma=1$.

\section{{\boldmath $\Lambda$} as a Function of Time}
\label{sec:t}

\subsection{Interpretation of the Time Co-ordinate}
\label{sec:time}

We begin with the oldest, and probably the most straightforward implementation
of the variable-$\Lambda$ idea; namely, one in which the cosmological
term is a simple power-law function of time, as set out
in eq.~(\ref{t-Law}):
\begin{equation}
\Lambda = {\cal A} \, \tau^{-\ell} ,
\label{Atell}
\end{equation}
where, for later convenience, we measure time in units of Hubble times
($\tau \equiv H_0 \, t$).
The case $\ell=2$ has previously been considered by several authors.
Each, however, has imposed supplementary conditions.
End\={o}, Fukui and others \cite{end77,ber86} operate 
in the context of a modified Brans-Dicke theory.  
Canuto {\em et al\/} \cite{can77} assume invariance under changes of scale,
while Lau \cite{lau85} adopts the Dirac large-number hypothesis 
(with a time-varying gravitational ``constant'' $G$) from the outset.
Berman \cite{ber91} requires that the density $\rho$ of 
ordinary matter {\em also\/} scale as $\tau^{-2}$, 
and that the deceleration parameter $q$ be a constant.
Beesham \cite{bee94} restricts his treatment to Bianchi Type~I models
with variable $G$.
Lopez and Nanopoulos \cite{lop96} take $\Lambda$ to have the same dependence
on the {\em scale factor\/} ($\Lambda \propto a^{-2}$) as on time,
for late times at least.
(These latter authors also make the important claim that a $\Lambda$-decay
ansatz of this kind could follow from certain versions of string theory.)
The question of the initial singularity is not addressed in
any of these papers.  In this paper, we examine the properties of models 
with the form~(\ref{Atell}) in a more comprehensive way.

Several conceptual issues should be dealt with before we proceed.
Firstly, since we are interested in oscillating as well as traditional
``big bang'' models, we are obliged to broaden the conventional
definition of cosmic time \cite{Cosmic}, in which it is set
to zero at the moment when $a=0$.  In those cases where the scale factor
never vanishes, we choose here to measure $\tau$ instead from the most recent
moment when $da/d\tau=0$ (the ``big bounce'').
In either case we refer to this as the ``initial moment'' and denote
it by $\tau=\tau_{\ast}\equiv 0.$  (Here and elsewhere, the subscript
``$\ast$'' will indicate quantities taken at this time.)

Secondly, eq.~(\ref{Atell}) implies that
$\Lambda\rightarrow\infty$ in the limit $\tau\rightarrow 0$, which may
not be realistic.  $\Lambda$ may go to infinity at some other
time ($\tau_{\infty}$, say) than the initial moment.  Or, more likely, its
divergent behavior may be cut off at some critical temperature
(at time $\tau_c$, say) by a phase transition or similar mechanism,
above which $\Lambda$ is effectively constant \cite{kaz80}.
So a more plausible formulation of eq.~(\ref{Atell}) might take the form:
\begin{equation}
\Lambda = \left\{ \begin{array}{ll}
                  \Lambda_c & \mbox{ when $\tau < \tau_c$ } \\
                  {\cal A} \, (\tau-\tau_{\infty})^{-\ell} & \mbox{ when 
                                          $\tau\geq\tau_c$ }
          \end{array} \right. ,
\label{Asplit}
\end{equation}
where continuity across the cutoff time $\tau_c$ implies that
$\Lambda_c \equiv {\cal A}\, (\tau_c-\tau_{\infty})^{-\ell}$.
However, the fact that the dynamical equations of \S\ref{sec:prelim} contain
no explicit time-dependence means that we can shift our time co-ordinate
via $(\tau-\tau_{\infty}) \rightarrow \tau$ with impunity.
The decay law~(\ref{Asplit}) then reverts to the
form~(\ref{Atell}), for all $\tau\geq\tau_c$ at least.
In practice we will assume that eq.~(\ref{Atell}) holds for all times 
of interest.  For earlier times the standard (Lema\^{\i}tre) solution applies,
with $\Lambda=\Lambda_c=$~constant.  If $\tau_c\approx\tau_{\infty}$,
then eq.~(\ref{Atell}) holds all the way to $\tau\approx 0$.

\subsection{Riccati's Equation}
\label{sec:Riccati}

We are now in a position to study the evolution of the scale factor.
Switching dependent variables from $a$ to $H$:
\begin{equation}
H \equiv \frac{\dot{a}}{a} = H_0 \left( \frac{da}{a \, d\tau} \right) ,
\label{Hdefn}
\end{equation}
we obtain for the differential equation~(\ref{mainDE}):
\begin{equation}
\frac{dH}{d\tau} = \left( \frac{-3\gamma}{2H_0} \right) H^2 +
                   \left( \frac{\gamma}{2H_0} \right) \Lambda +
                   \left( 1-\frac{3\gamma}{2} \right) \frac{k}{H_0 \, a^2} .
\label{1stDE}
\end{equation}
If we restrict ourselves to spatially flat universes ($k=0$), 
then the last term drops off, leaving a special case of Riccati's equation:
\begin{equation}
\frac{dH}{d\tau} = {\cal P}(\tau) \, H^2 + {\cal Q}(\tau) \, H + 
                   {\cal R}(\tau) ,
\label{RiccatiDE}
\end{equation}
where ${\cal P} \equiv - 3\gamma/2H_0$, ${\cal Q} \equiv 0$ and
${\cal R}(\tau) \equiv (\gamma/2H_0) \Lambda(\tau)$.
We will adopt this restriction for the remainder of \S\ref{sec:t}.
This has also been imposed in most of the special cases studied
so far \cite{end77,lau85,ber91,lop96}.

Solving Riccati's equation by standard methods \cite{DEs},
we change dependent variables from $H$ to $x$ via:
\begin{equation}
H \equiv -\frac{1}{{\cal P} x} \, \frac{dx}{d\tau} =
          \left( \frac{2H_0}{3\gamma} \right) \frac{dx}{x \, d\tau} ,
\label{xDefn}
\end{equation}
whereupon eq.~(\ref{1stDE}) takes the form [putting $k=0$ and 
inserting eq.~(\ref{Atell}) for $\Lambda(\tau)$]:
\begin{equation}
\tau^{\ell} \; \frac{d\,^2x}{d\tau^2} - \alpha \, x = 0 ,
\label{mainDE-t}
\end{equation}
with:
\begin{equation}
\alpha \equiv \frac{3\gamma^2 {\cal A}}{4H_0^2} .
\label{alphaDef1}
\end{equation}
This is now linear, as desired.
We will solve for $x(\tau)$ in the cases $\ell=1,2,3,4$.

Once $x(\tau)$ is found, the Hubble parameter $H(\tau)$ follows
immediately from the definition~(\ref{xDefn}).
Moreover, the scale factor $a(\tau)$ is also known,
as may be verified by comparing eqs.~(\ref{Hdefn}) and
(\ref{xDefn}) to yield:
\begin{equation}
a(\tau) = [ x(\tau) ]^{2/3\gamma} .
\label{xTOa}
\end{equation}
The constant $\alpha$ given by eq.~(\ref{alphaDef1})
can be fixed in terms of observational quantities as follows.
Applying the decay law~(\ref{Atell}) to the present epoch $\tau=\tau_0$,
and using the definition~(\ref{lamOm0defns}) of $\lambda_0$, we find that
${\cal A} = 3 H_0^2 \lambda_0 \tau_0^{\ell}$,
which can be substituted into eq.~(\ref{alphaDef1}) to yield:
\begin{equation}
\alpha = (3\gamma/2)^2 \lambda_0 \tau_0^{\ell} .
\label{alphaDefn}
\end{equation}
With the restriction $k=0$, eq.~(\ref{kDefn}) implies that $\Omega_0$ is 
given by $(1-\lambda_0)$ throughout \S\ref{sec:t}.

\subsection{The Case {\boldmath $\ell=1$}}
\label{sec:Casel=1}

We now proceed to the first case.
The differential equation~(\ref{mainDE-t}) reads:
\begin{equation}
\tau \; \frac{d\,^2x}{d\tau^2} - \alpha \, x = 0 ,
\label{mainDE-t-1}
\end{equation}
where $\alpha$ is given by eq.~(\ref{alphaDefn}):
\begin{equation}
\alpha = (3\gamma/2)^2 \lambda_0 \tau_0 .
\label{alphaDefn-1}
\end{equation}
Following standard techniques \cite{DEs}, we switch independent variables
from $\tau$ to $z \equiv 2\sqrt{-\alpha\tau}$, whereupon:
\begin{equation}
z^2 \, \frac{d\,^2x}{dz^2} - z \, \frac{dx}{dz} + z^2 \, x = 0 .
\label{mainDE-t-1b}
\end{equation}
This is transformable to Bessel's equation, with general solution
$x(z) = c_1 z J_1(z) + c_2 z Y_1 (z)$,
where $J_1(z)$ and $Y_1(z)$ are Bessel and Neumann functions of order one.
Eq.~(\ref{xTOa}) then gives for the scale factor:
\begin{equation}
a(\tau) = \tau^{1/3\gamma} \left[ c_1 J_1(z) + 
                                  c_2 Y_1 (z) \right]^{\; 2/3\gamma} ,
\label{aSoln-1}
\end{equation}
where we have absorbed a factor of $2\sqrt{-\alpha} \,$ into $c_1,c_2$.
The Hubble parameter is found by putting $x(z)$ into eq.~(\ref{xDefn}):
\begin{equation}
H(\tau) = H_0 \, \sqrt{-\lambda_0 \left( \frac{\tau_0}{\tau} \right)} \, 
                 \left[ \frac{c_1 J_0(z) + c_2 Y_0(z)}
                             {c_1 J_1(z) + c_2 Y_1(z)} \right] ,
\label{HSoln-1}
\end{equation}
where $J_0(z)$ and $Y_0(z)$ are Bessel and Neumann functions of order zero.
We note from the definition~(\ref{alphaDefn-1}) that $z(\tau)$, 
and hence $a(\tau)$ and $H(\tau)$, can only be real (for positive times)
if $\lambda_0 \leq 0$, which would imply a negative cosmological constant.
While this possibility has been considered in some contexts 
\cite{hoy97,NegLam}, we will see shortly that it leads in the present
theory to unrealistically short ages for the Universe.
Therefore the case $\ell=1$ is probably not realized in nature.

We can impose the following boundary conditions at the present epoch:
\begin{equation}
a(\tau_0) =a_0 \equiv 1 \; \; \; , \; \; \; H(\tau_0) = H_0 .
\label{BCs0}
\end{equation}
Substituting these expressions into eqs.~(\ref{aSoln-1}) and 
(\ref{HSoln-1}), it is straightforward to solve for $c_1$ and $c_2$:
\begin{eqnarray}
c_1 & = & \frac{\sqrt{-\lambda_0} \, Y_0(z_0) - Y_1(z_0)}
               {\sqrt{-\lambda_0 \tau_0} \, [ J_1(z_0) Y_0(z_0) -
               J_0(z_0) Y_1(z_0) ]}
\label{c1defn-1} \\
c_2 & = & \frac{-\sqrt{-\lambda_0} \, J_0(z_0) + J_1(z_0)}
               {\sqrt{-\lambda_0 \tau_0} \, [ J_1(z_0) Y_0(z_0) -
               J_0(z_0) Y_1(z_0) ]} ,
\label{c2defn-1}
\end{eqnarray}
where:
\begin{equation}
z_0 \equiv 3\gamma \tau_0 \sqrt{-\lambda_0} .
\label{z0defn}
\end{equation}
To keep $a(\tau)$ finite at $\tau =0$, we require $c_2=0$,
since $Y_1(z)$ diverges logarithmically at $z=0$.
This constitutes our third boundary condition.
In conjunction with eq.~(\ref{c2defn-1}) it implies that
$J_1(z_0) - \sqrt{-\lambda_0} \, J_0(z_0) = 0$.
This equation may be solved numerically for $\tau_0$ as a function 
of $\lambda_0$, with the help of the definition~(\ref{z0defn}).
The results can then be substituted back into eq.~(\ref{c1defn-1})
to fix the value of $c_1$.  With $c_1$ and $c_2$ both known,
$a(\tau)$ and $H(\tau)$ are given by eqs.~(\ref{aSoln-1}) and
(\ref{HSoln-1}) respectively.

The evolution of the scale factor for this case is illustrated in
Fig.~\ref{figure1} for various values of $\lambda_0$, assuming
$\gamma=1$ (after Felten and Isaacman \cite{fel86}).
In particular, we have followed these authors in plotting the
scale factor $a$ as a function of $(\tau-\tau_0)$, rather than $\tau$
for each curve.
This has the effect of shifting all the curves so that they
intersect at $(0,1)$, which marks the present epoch.
[Recall that we have chosen units such that $a_0=1$, eq.~(\ref{Useful}).]

We have plotted for four Hubble times into the future,
and one Hubble time into the past.  It may be seen
that the $\lambda_0=0$ curve (solid line) intersects the time axis
at $(\tau-\tau_0)=-2/3$, which confirms the well-known
rule that the age of a flat universe with no cosmological constant
is $\tau_0=2/3$.  The models with $\lambda_0<0$ (dashed lines) are 
younger than this, which considerably diminishes their attractiveness.
The $\lambda_0=-1$ model, for instance, has $\tau_0=0.48$, while
the $\lambda_0=-3$ model has $\tau_0=0.35$.
If we use a current widely-accepted value of
$H_0=73 \pm 10$~km~s$^{-1}$~Mpc$^{-1}$ \cite{fre96} for the Hubble
parameter, then (recalling that $t=\tau/H_0$) we see that the age 
of the Universe in these models can be no more than 7.4 and 
5.4~billion years old respectively.
This conflicts badly with estimated globular cluster ages, 
which are thought to be at least 9.6~billion years old in some cases
\cite{cha97}.  The situation improves slightly
if one switches to the lower values for $H_0$ which are reported by
some authors \cite{san96}.  If $H_0=55 \pm 10$~km~s$^{-1}$~Mpc$^{-1}$,
then the maximum possible age for these two
models increases to 10.4 and 7.6~billion years respectively.
One can safely rule out models with $\lambda_0 < -1$ on this basis,
while models in the range $-1 < \lambda_0 \leq 0$ remain
marginally viable at best.

Fig.~\ref{figure1} tells us that, while flat models with $\lambda_0=0$
will continue to expand indefinitely as usual, those with negative values 
of $\lambda_0$ will experience eventual recollapse.
This can be understood by looking at eq.~(\ref{accel}),
which shows that there are two contributions to the deceleration:
one (which goes as $-\rho a$) due to ordinary matter and the
other (which goes as $+\Lambda a$) due to the negative
cosmological constant.  Because the density $\rho$ of matter rapidly
thins out with expansion, the first contribution alone is not enough
to close the Universe when $k=0$.  The second contribution, however,
is diluted much more slowly ($\Lambda$ drops off as only
$\tau^{-1}$ in this case), and is therefore sufficient to turn the expansion
around eventually, no matter how small its value at the present time.
Thus, models with $\lambda_0=-1$ and $\lambda_0=-3$ encounter the 
``big crunch'' after only 2.94 and 1.21~Hubble times respectively.

\subsection{The Case {\boldmath $\ell=2$}}

For this case, eq.~(\ref{mainDE-t}) takes the form:
\begin{equation}
\tau^2 \; \frac{d\,^2x}{d\tau^2} - \alpha \, x = 0 ,
\label{mainDE-t-2}
\end{equation}
with $\alpha$ given by eq.~(\ref{alphaDefn}) as follows:
\begin{equation}
\alpha = (3\gamma/2)^2 \lambda_0 \tau_0^2 .
\label{alphaDefn-2}
\end{equation}
This is a special case of Euler's differential equation.
Applying standard methods \cite{DEs}, we switch independent variables
via $y \equiv \ln\tau$ to recast eq.~(\ref{mainDE-t-2}) in the form:
\begin{equation}
\frac{d\,^2x}{dy^2} - \frac{dx}{dy} - \alpha \, x = 0 .
\label{EulerDE}
\end{equation}
This now has constant coefficients, as desired.
There are three subcases, according as $\lambda_0$ is greater than, 
equal to, or less than $-1/(3\gamma\tau_0)^2$.

\subsubsection{The Subcase $\lambda_0 > -1/(3\gamma\tau_0)^2$}

Since we expect on observational grounds that $\lambda_0$ is probably 
positive, this is the most physical of the subcases.
Solution of eq.~(\ref{EulerDE}) for $x(y)$ and hence $x(\tau)$ is 
straightforward.  The scale factor and Hubble parameter are found 
from eqs.~(\ref{xTOa}) and (\ref{xDefn}) to be:
\begin{eqnarray}
a(\tau) & = & \tau^{1/3\gamma} \, \left( c_1 \tau^{m_0} + c_2 
                                  \tau^{-m_0} \right)^{2/3\gamma}
\label{aSoln-2a-2} \\
H(\tau) & = & \frac{2H_0}{3\gamma} \, \left[ \frac
              {m_1 c_1 \tau^{m_0} + m_2 c_2 \tau^{m_0}}
              {\tau \, ( c_1 \tau^{m_0} + c_2 \tau^{-m_0})} \right] ,
\label{HSoln-2a-2}
\end{eqnarray}
where $m_0 \equiv \frac{1}{2} \sqrt{1+(3\gamma\tau_0)^2 \lambda_0}$.
It is clear that $a(\tau)$ diverges at $\tau=0$ for $\lambda_0>0$ unless
$c_2 = 0$, which we consequently adopt as a boundary condition
(as in the $\ell=1$ case).  Eqs.~(\ref{aSoln-2a-2}) and (\ref{HSoln-2a-2}) 
simplify to:
\begin{eqnarray}
a(\tau) & = & (c_1 \tau^{m_1}\,)^{2/3\gamma}
\label{aSoln-2a-3} \\
H(\tau) & = & \frac{2H_0}{3\gamma} \, \left[ \frac{m_1}{\tau} \right] ,
\label{HSoln-2a-3}
\end{eqnarray}
where $m_1 \equiv 1/2+m_0$.
We then apply the boundary conditions~(\ref{BCs0}) at $\tau=\tau_0$.
Eq.~(\ref{HSoln-2a-3}) with $H(\tau_0)=H_0$ implies that:
\begin{equation}
\tau_0 = 2m_1/3\gamma . 
\label{Age-2a-1}
\end{equation}
Substituting this result into eq.~(\ref{aSoln-2a-3}), with
$a(\tau_0)=1$, we find that $c_1 = (3\gamma/2m_1)^{m_1}$,
which can be put back into eq.~(\ref{aSoln-2a-3}) to yield
the following expression for $a(\tau)$:
\begin{equation}
a(\tau) = \left( \frac{\tau}{\tau_0} \right)^{2m_1/3\gamma} .
\label{aSoln-2a-4}
\end{equation}
The scale factor expands as a simple power-law function of time.
This is consistent with previous special cases obtained for $\ell=2$:
End\={o} and Fukui's $a(\tau) \propto \tau^{2n/3(n-1)}$ \cite{end77},
Berman and Som's $a(\tau) \propto \tau^{1/m}$ \cite{ber86},
Lau and Beesham's $a(\tau) \propto \tau^{1/3}$ \cite{lau85,bee94},
Berman's $a(\tau) \propto \tau^{2/3}$ \cite{ber91}, and
Lopez and Nanopoulos' $a(\tau) \propto \tau$ \cite{lop96}.

In conjunction with the definitions of $m_0$ and $m_1$, 
eq.~(\ref{Age-2a-1}) fixes the age of the Universe at:
\begin{equation}
\tau_0 = \frac{2}{3\gamma(1-\lambda_0)} , 
\label{Age-2a-2}
\end{equation}
from which we draw two conclusions:  firstly, that all models satisfying 
the boundary conditions obey $\lambda_0 < 1$;
and secondly, that the age of the Universe $\tau_0 \rightarrow \infty$
as $\lambda_0 \rightarrow 1$.  The initial singularity can thus be
pushed back arbitrarily far into the past.
We also find a lower limit on the age of the Universe in these
models by noting that $m_1=(m_0+1/2)>1/2$.  Inserting this into
eq.~(\ref{Age-2a-1}) produces the result:
\begin{equation}
\tau_0 > 1/3\gamma . 
\label{Age-2a-3}
\end{equation}
In other words, assuming zero pressure ($\gamma=1$), all models 
in this case have survived for at least one-third of a Hubble time.  
Adopting a recent observational upper limit of 83~km~s$^{-1}$~Mpc$^{-1}$ 
\cite{fre96}, this implies a minimum age of at least 3.9~billion years.
Finally, putting eq.~(\ref{Age-2a-3}) back into the
expression~(\ref{Age-2a-2}) for $\tau_0$, we find that
$\lambda_0 > -1$, which fixes the critical value of $\lambda_0$ 
separating this subcase from the other two.

\subsubsection{The Subcase $\lambda_0 = -1/(3\gamma\tau_0)^2$}

Solution of eq.~(\ref{EulerDE}) is also straightforward, 
and one finds from eqs.~(\ref{xTOa}) and (\ref{xDefn}) the
following general expressions for $a(\tau)$ and $H(\tau)$:
\begin{eqnarray}
a(\tau) & = & \tau^{1/3\gamma} \, (c_3 + c_4 \ln \tau)^{2/3\gamma}
\label{aSoln-2b-1} \\
H(\tau) & = & \frac{2H_0}{3\gamma} \, \left[ \frac
              {(c_3+2c_4) + c_4 \ln\tau}
              {2\tau ( c_3 + c_4 \ln\tau)} \right] ,
\label{HSoln-2b-1}
\end{eqnarray}
where $c_3,c_4$ are arbitrary constants.
To keep $a(\tau)$ finite at $\tau=0$ we require $c_4=0$, so that:
\begin{eqnarray}
a(\tau) & = & (c_3 \sqrt{\tau})^{2/3\gamma}
\label{aSoln-2b-2} \\
H(\tau) & = & \frac{H_0}{3\gamma\tau} .
\label{HSoln-2b-2}
\end{eqnarray}
Inserting $H(\tau_0)=H_0$ into eq.~(\ref{HSoln-2b-2}), we find
for the age of the Universe in this model:
\begin{equation}
\tau_0 = 1/3\gamma , 
\label{Age-2b}
\end{equation}
which is exactly the limit $\lambda_0 \rightarrow -1$ in 
eq.~(\ref{Age-2a-2}).  Eq.~(\ref{Age-2b}) also corresponds to the 
lower limit allowed by eq.~(\ref{Age-2a-3}), as one might expect.

Substituting the age~(\ref{Age-2b}) into eq.~(\ref{aSoln-2b-2}),
meanwhile, and imposing $a(\tau_0)=1$ as usual, we find that
$c_3 = \sqrt{3\gamma}$.  Inserted back into eq.~(\ref{aSoln-2b-2}), 
this yields:
\begin{equation}
a(\tau) = \left( \frac{\tau}{\tau_0} \right)^{1/3\gamma} ,
\end{equation}
which joins smoothly onto the result~(\ref{aSoln-2a-4}) from the
previous subcase.

\subsubsection{The Subcase $\lambda_0 < -1/(3\gamma\tau_0)^2$}
\label{sec:deSitter}

Solution of eq.~(\ref{EulerDE}) is also straightforward and leads 
to the following general expressions for $a(\tau)$ and $H(\tau)$:
\begin{eqnarray}
a(\tau) & = & \tau^{1/3\gamma} \, \left[ c_5 \sin (m_3 \ln\tau) + 
              c_6 \cos (m_3 \ln\tau) \right]^{2/3\gamma}
\label{aSoln-2c-1} \\
H(\tau) & = & \frac{2H_0}{3\gamma} \, \left\{ \frac
              {(c_5-2m_3c_6)\sin(m_3\ln\tau)+(c_6+2m_3c_5)\cos(m_3\ln\tau)}
              {2\tau[c_5\sin(m_3\ln\tau)+c_6\cos(m_3\ln\tau)]} \right\} ,
\label{HSoln-2c-1}
\end{eqnarray}
where $m_3 \equiv \frac{1}{2} \sqrt{-(3\gamma\tau_0)^2 \lambda_0 - 1}$
and $c_5,c_6$ are arbitrary constants.
Application of the boundary conditions~(\ref{BCs0}) gives
$c_5$ and $c_6$ in terms of $\tau_0$:
\begin{eqnarray}
c_5 & = & \frac{1}{\sqrt{\tau_0}} \sin(m_3\ln\tau_0) + \frac{1}{m_3}
          \left[ \left( \frac{3\gamma}{2} \right) \sqrt{\tau_0} -
          \frac{1}{2\sqrt{\tau_0}} \right] \cos(m_3\ln\tau_0) \\
\label{c5defn-2}
c_6 & = & \frac{1}{\sqrt{\tau_0}} \cos(m_3\ln\tau_0) - \frac{1}{m_3}
          \left[ \left( \frac{3\gamma}{2} \right) \sqrt{\tau_0} -
          \frac{1}{2\sqrt{\tau_0}} \right] \sin(m_3\ln\tau_0) .
\label{c6defn-2}
\end{eqnarray}
As usual, we need a third boundary condition to fix the value of $\tau_0$.
Unlike the previous two subcases, we cannot keep $a(\tau)$ finite at
$\tau=0$ by setting one of $c_5,c_6$ to zero.
Instead we have adopted a numerical approach, searching iteratively
for the value of $\tau_0$ consistent with the boundary conditions 
(ie., with the requirement that either $a(\tau)$ or $H(\tau)$ go smoothly 
to zero as $\tau\rightarrow 0$).  The values of $c_5$ and $c_6$ then follow 
from eqs.~(\ref{c5defn-2}) and (\ref{c6defn-2}) respectively.

The evolution of the scale factor for this case is illustrated in
Fig.~\ref{figure2}, which has exactly the same format as Fig.~\ref{figure1},
except that we have plotted for three Hubble times into the past
instead of one.
Fig.~\ref{figure2} exhibits a richer variety of solutions
than Fig.~\ref{figure1}.  The most noticeable difference is the existence
of solutions for positive $\lambda_0$ (short-dashed lines).  Of particular
interest is the limiting case $\lambda_0=1$, which only approaches
$a=0$ asymptotically as $\tau \rightarrow -\infty$.
This case is not very realistic, however, as it has zero density
(since $\Omega_0=1-\lambda_0$).
It is in fact the empty de~Sitter model.  The same solution is found
for these values of $\lambda_0$ and $\Omega_0$ in conventional
Lema\^{\i}tre cosmology with $\Lambda=$~constant \cite{fel86}.

Fig.~\ref{figure2} therefore shows that we cannot avoid the big bang
in a theory with $\Lambda \propto \tau^{-2}$ and $k=0$.
We can, however, significantly extend the age of the Universe.
Suppose we choose $\lambda_0=0.5$, for example; a value compatible
with the tightest observational bounds thus far \cite{per97}.
Fig.~\ref{figure2} shows that this model would have come into 
being 1.33~Hubble times ago [see also eq.~(\ref{Age-2a-2}) above].
Even if we adopt the upper limit of 83~km~s$^{-1}$~Mpc$^{-1}$
for $H_0$ \cite{fre96}, this translates into
an age of 15.7~billion years --- more than enough time for the oldest 
globular clusters to form \cite{cha97}.
By way of comparison, a {\em constant\/}-$\Lambda$ model with 
$\lambda_0=\Omega_0=0.5$ has an age of only 0.83~Hubble times.

Flat models with no cosmological constant are
represented in Fig.~\ref{figure2} by the curve labelled
$\lambda_0=0$ (solid line).  As in the $\ell=1$ case, these have
an age of 2/3~Hubble times.  Models with a negative $\lambda_0$ 
(long-dashed lines) all have shorter ages, as in the $\ell=1$ case.
The difference is that they are now {\em even shorter\/},
because the negative cosmological term is driven to high negative
values more quickly in the past direction when $\ell=2$.
Thus, the age of the $\lambda_0=-1$ model has dropped from 0.48 to 
0.33~Hubble times, while that of the $\lambda_0=-3$ model is now
only $\tau_0=0.30$, down from $0.35$ in Fig.~\ref{figure1}.

Fig.~\ref{figure2} indicates that the $\lambda_0<0$ models tend toward 
eventual recollapse, as they did in the $\ell=1$ case.
However, this process now takes much longer.  In other words, 
while the $\ell=2$ models are younger, their life expectancies
are considerably greater.  This can be understood 
by means of the same argument as before (\S\ref{sec:Casel=1}).
The larger value of $\ell$ means that the contribution of
the cosmological term to the deceleration drops off more quickly
in the future direction,
thereby postponing recollapse for a longer period of time.
Thus, models with $\lambda_0=-1$ and $\lambda_0=-3$ now survive for
$t \gg 5$ and $t=3.37$~Hubble times respectively.

\subsection{The Case {\boldmath $\ell=3$}}

For this case, eq.~(\ref{mainDE-t}) reads:
\begin{equation}
\tau^3 \; \frac{d\,^2x}{d\tau^2} - \alpha \, x = 0 ,
\label{mainDE-t-3}
\end{equation}
where $\alpha$ is given by eq.~(\ref{alphaDefn}):
\begin{equation}
\alpha = (3\gamma/2)^2 \lambda_0 \tau_0^3 .
\label{alphaDefn-3}
\end{equation}
Following standard techniques \cite{DEs}, we switch independent variables 
from $\tau$ to $z \equiv 1/\tau$, whereupon:
\begin{equation}
z \, \frac{d\,^2x}{dz^2} + 2 \, \frac{dx}{dz} - \alpha \, x = 0 .
\label{mainDE-t-3b}
\end{equation}
This is now in a similar form to eq.~(\ref{mainDE-t-1}) in the $\ell=1$
case, and it can be solved in the same manner \cite{DEs}.
Changing independent variables again, from $z$ to
$y \equiv 2\sqrt{-\alpha z}$, eq.~(\ref{mainDE-t-3b}) takes the form:
\begin{equation}
y^2 \, \frac{d\,^2x}{dy^2} + 3 y \, \frac{dx}{dy} + y^2 \, x = 0 .
\end{equation}
This is again transformable to Bessel's equation, but with
a different general solution
$x(y) = y^{-1} [ c_1 J_1(y) + c_2 Y_1(y) ]$,
where $c_1,c_2$ are arbitrary constants.
Eq.~(\ref{xTOa}) then gives $a(\tau)$, as usual:
\begin{equation}
a(\tau) = \tau^{1/3\gamma} \left[ c_1 J_1(y) +
                                  c_2 Y_1 (y) \right]^{2/3\gamma} ,
\label{aSoln-3}
\end{equation}
where $y(\tau) = 2\sqrt{-\alpha/\tau}$ and we have
absorbed a factor of $2\sqrt{-\alpha}$ into $c_1,c_2$.
The Hubble parameter is found as usual by putting $x(y)$ into 
eq.~(\ref{xDefn}):
\begin{equation}
H(\tau) = H_0 \, \sqrt{-\lambda_0 \left( \frac{\tau_0}{\tau} \right)^3} \,
                 \left[ \frac{c_1 J_2(y) + c_2 Y_2(y)}
                             {c_1 J_1(y) + c_2 Y_1(y)} \right] ,
\label{HSoln-3}
\end{equation}
where $J_2(y)$ and $Y_2(y)$ are Bessel and Neumann functions of 
order two.  As in the $\ell=1$ case, these solutions are real-valued
(for $\tau > 0$) only if the cosmological term is negative.

In conjunction with the boundary conditions~(\ref{BCs0}),
eqs.~(\ref{aSoln-3}) and (\ref{HSoln-3}) give for $c_1$ and $c_2$:
\begin{eqnarray}
c_1 & = & \frac{\sqrt{-\lambda_0} \, Y_2(y_0) - Y_1(y_0)}
               {\sqrt{-\lambda_0 \tau_0} \, [ J_1(y_0) Y_2(y_0) -
               J_2(y_0) Y_1(y_0) ]}
\label{c1defn-3} \\
c_2 & = & \frac{-\sqrt{-\lambda_0} \, J_2(y_0) + J_1(y_0)}
               {\sqrt{-\lambda_0 \tau_0} \, [ J_1(y_0) Y_2(y_0) -
               J_2(y_0) Y_1(y_0) ]} ,
\label{c2defn-3}
\end{eqnarray}
where $y_0=z_0$ is given by eq.~(\ref{z0defn}).
We require one additional boundary condition to fix $\tau_0$.
Unfortunately, as in the previous subcase, the
procedure is complicated by the fact that both terms in
eq.~(\ref{aSoln-3}) diverge at $\tau=0$, whereas we expect that 
the scale factor as a whole should behave smoothly there.

We can make this more precise by employing the asymptotic expressions
for $J_1(y)$ and $Y_1(y)$ at large $y$ (ie., small $\tau$).
We find (for $\tau \ll 1$):
\begin{equation}
a(\tau) \approx \tau^{1/2\gamma} \left[
                C_1 \sin\left(\frac{\omega_0}{\sqrt{\tau}}\right) +
                C_2 \cos\left(\frac{\omega_0}{\sqrt{\tau}}\right)
                                 \right]^{2/3\gamma} ,
\label{aSoln-3b}
\end{equation}
where $C_1 \equiv C_0^{2/3\gamma}(c_1-c_2)$,
$C_2 \equiv -C_0^{2/3\gamma}(c_1+c_2)$,
$C_0 \equiv 1/\sqrt{\pi\omega_0}$ and
$\omega_0 \equiv 3\gamma\tau_0 \sqrt{-\lambda_0\tau_0}$.
This goes smoothly to zero for $\tau\rightarrow 0$.
In order to find the correct (ie., self-consistent) value of $\tau_0$,
we use the same numerical approach as in the $\ell=2$ case.
Once $\tau_0$ is obtained, the values of $c_1$ and $c_2$ follow from 
eqs.~(\ref{c1defn-3}) and (\ref{c2defn-3}) respectively.

The evolution of the scale factor for this case is illustrated in
Fig.~\ref{figure3}, which has exactly the same format as Fig.~\ref{figure1}.
The $\lambda_0=0$ model (solid line) has an age of 2/3~Hubble times,
while those with negative values of $\lambda_0$ (long-dashed lines) all have 
shorter ages, as usual.  The age of the $\lambda_0=-3$ model, for example,
has dropped from 0.30 to just 0.27~Hubble times.
It is unlikely that any of these models could describe the
real universe, given the observational constraints on $H_0$ and
$\tau_0$ (\S\ref{sec:Casel=1}).

The main difference between Fig.~\ref{figure3} and its predecessors 
occurs at large times, where we observe that the curves all straighten out,
and show no sign of leading to a recollapse of the scale factor.
The explanation for this is that the (negative) cosmological term 
is now decaying so quickly with time that, like ordinary matter, 
it is no longer sufficient to turn the expansion around.

\subsection{The Case {\boldmath $\ell=4$}}

For this case, eq.~(\ref{mainDE-t}) reads:
\begin{equation}
\tau^4 \; \frac{d\,^2x}{d\tau^2} - \alpha \, x = 0 ,
\label{mainDE-t-4}
\end{equation}
with $\alpha$ given by eq.~(\ref{alphaDefn}) as follows:
\begin{equation}
\alpha = (3\gamma/2)^2 \lambda_0 \tau_0^4 .
\label{alphaDefn-4}
\end{equation}
Switching independent variables via $z\equiv 1/\tau$ as in the previous
case, we find that eq.~(\ref{mainDE-t-4}) takes the form:
\begin{equation}
z \, \frac{d\,^2x}{dz^2} + 2 \, \frac{dx}{dz} - \alpha \, z \, x = 0 .
\label{mainDE-t-4b}
\end{equation}
Employing standard methods \cite{DEs}, we switch dependent
variables from $z$ to $y$ via $dy/dz \equiv z \, x(z)$.
It may then be verified that $y(z)$ satisfies the familiar equation:
\begin{equation}
\frac{d\,^2y}{dz^2} = \alpha \, y .
\label{SHM}
\end{equation}
The solutions of eq.~(\ref{SHM}) are well-known; there are three subcases
to consider, according as $\alpha$ [and hence $\lambda_0$, 
eq.~(\ref{alphaDefn-4})] is positive, zero, or negative.

\subsubsection{The Subcase $\lambda_0 > 0$}

Since we expect on observational grounds that $\lambda_0$ is probably
positive, this is the most physical of the three.
Solution of eq.~(\ref{SHM}) for $y(z)$ and hence $x(z)$ is
straightforward.  The scale factor and Hubble parameter are
given by eqs.~(\ref{xTOa}) and (\ref{xDefn}):
\begin{eqnarray}
a(\tau) & = & \tau^{2/3\gamma} \, \left[ 
              c_1 \exp \left( \frac{\sqrt{\alpha}}{\tau} \right) + c_2 \exp 
              \left( \frac{-\sqrt{\alpha}}{\tau} \right) \right]^{2/3\gamma}
\label{aSoln-4a-1} \\
H(\tau) & = & \frac{2H_0}{3\gamma} \, \left[ \frac
              { c_1 (1-\sqrt{\alpha}/\tau)\exp(\sqrt{\alpha}/\tau) +
                c_2 (1+\sqrt{\alpha}/\tau)\exp(-\sqrt{\alpha}/\tau) }
              { c_1 \tau \exp(\sqrt{\alpha}/\tau) +
                c_2 \tau \exp(-\sqrt{\alpha}/\tau) } \right] ,
\label{HSoln-4a-1}
\end{eqnarray}
where $c_1,c_2$ are arbitrary constants.
It is clear that $a(\tau)$ diverges at $\tau=0$ unless
$c_1 = 0$, which we consequently assume.
Eqs.~(\ref{aSoln-4a-1}) and (\ref{HSoln-4a-1}) simplify:
\begin{eqnarray}
a(\tau) & = & \left[ c_2 \tau \exp \left( \frac{-\sqrt{\alpha}}{\tau}
              \right) \right]^{2/3\gamma} 
\label{aSoln-4a-2} \\
H(\tau) & = & \frac{2H_0}{3\gamma} \, \left( 1 + \frac{\sqrt{\alpha}}{\tau}
              \right) \, \frac{1}{\tau} . 
\label{HSoln-4a-2}
\end{eqnarray}
We then apply the boundary conditions~(\ref{BCs0}) at $\tau=\tau_0$,
as usual.  Eq.~(\ref{HSoln-4a-2}) with $H(\tau_0)=H_0$ fixes the age at:
\begin{equation}
\tau_0 = \frac{2}{3\gamma(1-\sqrt{\lambda_0})} .  
\label{Age-4a}
\end{equation}
Substituting this result into eq.~(\ref{aSoln-4a-2}) with
$a(\tau_0)=1$, we find that $c_2 = (1/\tau_0) \, \exp (\sqrt{\alpha}/\tau_0)$,
which can be put back into eq.~(\ref{aSoln-4a-2}) to yield this
expression for $a(\tau)$:
\begin{equation}
a(\tau) = \left[ \left( \frac{\tau}{\tau_0} \right) \exp \left(
          \frac{\sqrt{\alpha}}{\tau_0} - \frac{\sqrt{\alpha}}{\tau}
          \right) \right]^{2/3\gamma} .
\label{aSoln-4a-3}
\end{equation}
We can draw a number of conclusions from eq.~(\ref{Age-4a}):
firstly, that $\lambda_0 < 1$;
and secondly, that the age of the Universe $\tau_0 \rightarrow \infty$
as $\lambda_0 \rightarrow 1$.  This is reminiscent of the $\ell=2$ case,
and in fact eq.~(\ref{Age-4a}) is almost identical to eq.~(\ref{Age-2a-1}),
the only difference being that $\lambda_0$ in the denominator has been
replaced by $\sqrt{\lambda_0}$.  Therefore, for the same value of
$\lambda_0$, the $\ell=4$ models are longer-lived by a factor of 
$(1-\lambda_0)/(1-\sqrt{\lambda_0})$.  This is due to the fact that, 
for positive $\Lambda$, the higher value of $\ell$ means that
the cosmological term exerts a greater repulsive force in the past
direction, and is consequently able to push the big bang back more 
effectively.  As in the $\ell=2$ case, we also find a lower limit on 
the age of these models.  This is, from eq.~(\ref{Age-4a}):
\begin{equation}
\tau_0 > 2/3\gamma .
\end{equation}
This is twice as long as in the $\ell=2$ case, but here
the reason is simply that this lower limit corresponds to the 
case $\lambda_0=0$ (not $\lambda_0=-1$ as before).
With dust-like conditions ($\gamma=1$) and the upper bound
$H_0 \leq 83$~km~s$^{-1}$~Mpc$^{-1}$ \cite{fre96},
we now find a minimum age of at least 7.9~billion years.

\subsubsection{The Subcase $\lambda_0 = 0$}

For this subcase eq.~(\ref{SHM}) is trivial.  Using eqs.~(\ref{xTOa}) 
and (\ref{xDefn}) we find immediately that:
\begin{eqnarray}
a(\tau) & = & (c_3 + c_4 \tau)^{2/3\gamma} 
\label{aSoln-4b} \\
H(\tau) & = & \frac{2H_0}{3\gamma} \, \left( \frac{c_4}{c_3+c_4 \tau}
              \right) 
\label{HSoln-4b} ,
\end{eqnarray}
where $c_3,c_4$ are arbitrary constants.
For the first time we have a scale factor with the potential 
to go smoothly to some finite value {\em other than\/}
zero at $\tau=0$.  Let us pursue this possibility and see if a nonsingular
solution is possible.  Instead of the boundary condition $a(\tau_0)=1$, 
we impose $a(0)=a_{\ast}$,
where $a_{\ast}$ is the minimum value of the scale factor.
In eq.~(\ref{aSoln-4b}) this implies that
$c_3 = a_{\ast}^{3\gamma/2}$.
Inserting this result back into eq.~(\ref{aSoln-4b}) and applying
the usual condition $a(\tau_0)=1$, we obtain $c_4 = (1-c_3)/\tau_0$.
Substituting this into eq.~(\ref{HSoln-4b}) and applying the third
boundary condition $H(\tau_0)=H_0$, we obtain for the age of the
Universe:
\begin{equation}
\tau_0 = 2(1-c_3)/3\gamma .
\label{Age-4b}
\end{equation}
This result matches onto that of the previous subcase, eq.~(\ref{Age-4a}), 
only if $c_3 = 0$, which also implies that $a_{\ast}=0$.
The present case is therefore singular at $\tau=0$, like all the others 
studied in this section.  The age of the Universe is given by 
eq.~(\ref{Age-4b}) as $\tau_0 = 2/3\gamma$.
Putting these results back into eq.~(\ref{aSoln-4b}), we find:
\begin{equation}
a(\tau) = \left( \frac{\tau}{\tau_0} \right)^{2/3\gamma} .
\end{equation}
This is just the standard $k=0$ solution with no cosmological term,
as might have been expected.

\subsubsection{The Subcase $\lambda_0 < 0$}

Solution of eq.~(\ref{SHM}) is again straightforward and leads
via eqs.~(\ref{xTOa}) and (\ref{xDefn}) to:
\begin{eqnarray}
a(\tau) & = & \tau^{2/3\gamma} \, \left[
              c_5 \sin \left( \frac{\sqrt{-\alpha}}{\tau} \right) +
              c_6 \cos \left( \frac{\sqrt{-\alpha}}{\tau} \right)
              \right]^{2/3\gamma} 
\label{aSoln-4c-1} \\
H(\tau) & = & \frac{2H_0}{3\gamma} \, \left\{ \frac
              {[c_5+(\sqrt{-\alpha}/\tau)c_6]\sin(\sqrt{-\alpha}/\tau) +
               [c_6-(\sqrt{-\alpha}/\tau)c_5]\cos(\sqrt{-\alpha}/\tau)}
              {c_5\tau\sin(\sqrt{-\alpha}/\tau) +
               c_6\tau\cos(\sqrt{-\alpha}/\tau)} \right\} ,
              \nonumber
\end{eqnarray}
where $c_5,c_6$ are arbitrary constants.
Application of the boundary conditions~(\ref{BCs0}) fixes
these constants in terms of $\tau_0$:
\begin{eqnarray}
c_5 & = & \frac{1}{\sqrt{-\alpha}} \, \left[ \beta_0 \sin\beta_0 +
          \cos\beta_0 \left( 1 - \frac{3\gamma}{2} \, \tau_0 \right)
          \right] 
\label{c5defn-4} \\
c_6 & = & \frac{1}{\sqrt{-\alpha}} \, \left[ \beta_0 \cos\beta_0 -
          \sin\beta_0 \left( 1 - \frac{3\gamma}{2} \, \tau_0 \right)
          \right] 
\label{c6defn-4} ,
\end{eqnarray}
where 
$\beta_0 \equiv \left( \frac{3\gamma}{2} \right) \tau_0 \sqrt{-\lambda_0}$.
As usual, we require one additional boundary condition to fix the value
of $\tau_0$.  The situation is again complicated by the fact that
both terms in eq.~(\ref{aSoln-4c-1}) diverge at $\tau=0$, whereas $a(\tau)$
itself goes smoothly to zero there.  [In fact, eq.~(\ref{aSoln-4c-1}) 
has exactly the same form as the asymptotic expression~(\ref{aSoln-3b}) 
in the $\ell=3$ case.]
We therefore have recourse once again to the numerical method described
in \S\ref{sec:deSitter}.
Once $\tau_0$ is obtained in this way, the values of $c_5$ and $c_6$
are fixed by eqs.~(\ref{c5defn-4}) and (\ref{c6defn-4}).

The evolution of the scale factor for this case is illustrated in
Fig.~\ref{figure4}, which has exactly the same format as Fig.~\ref{figure2}.
Several features may be noted.  To begin with, we see that models with 
$\Lambda \propto \tau^{-4}$ are qualitatively the same as those with 
$\Lambda \propto \tau^{-2}$ for positive $\lambda_0$, 
and qualitatively similar to those with $\Lambda \propto \tau^{-3}$ 
for negative $\lambda_0$.

There are important quantitative differences, however.
Models with positive $\lambda_0$ are significantly older.
With $\lambda_0=0.5$, for example, $\tau_0$ is now 2.28~Hubble times --- 
older by a factor of 1.71~times than the equivalent $\ell=2$ model, 
exactly as predicted in the discussion following eq.~(\ref{aSoln-4a-3}).
This is once again due to the fact that, with $\ell=4$, a positive
$\Lambda$-term increases in size very rapidly in the past direction.
Negative-$\lambda_0$ models, on the other hand, are once again 
younger.  The age of the $\lambda_0=-3$ model, for instance,
has dropped from 0.27 to just 0.25~Hubble times.  
And finally, in the future direction, we see that there is no longer very
much distinction between the $\lambda_0<0$ and $\lambda_0=0$ models,
compared to Fig.~\ref{figure3}.  The contribution of the 
cosmological term to the deceleration of the scale factor now drops off
so quickly that it rapidly becomes irrelevant.

\section{{\boldmath $\Lambda$} as a Function of the Scale Factor}
\label{sec:a}

\subsection{Previous Work}
\label{sec:prev}

We now move on to consider decay laws of the form set out in 
eq.~(\ref{a-Law}):
\begin{equation}
\Lambda = {\cal B} \, a^{-m} .
\label{Bam}
\end{equation}
The scale factor may be more natural than time as an independent variable,
to the extent that many physical quantities (such as temperature) depend 
more simply on $a$ than $t$.
Nearly half of the decay laws listed in Table~\ref{table1} contain
terms of the form~(\ref{Bam}).
The case $m=2$ has been singled out for the most attention 
\cite{lop96,oze86,che90,cal92}, and is motivated by some
dimensional arguments \cite{lop96,che90}.
A second group of authors has concentrated on values of $m \approx 4$ 
\cite{fre87,gas87,sat90,ove93}, for which the $\Lambda$-term behaves 
like ordinary radiation.  It has been shown that, for certain kinds of
$\Lambda$-decay, the lower of these two values of $m$ is 
thermodynamically more stable \cite{pav91}.
A third idea, that the cosmological term scales with $a$ like ordinary
matter ($m=3$), follows from one interpretation of an intriguing
new scale-invariant extension of general relativity \cite{hoy97}.

There are also some studies in which the value of $m$ is not fixed 
{\em a priori\/}.  Ages of general-$m$ models have been calculated
and agree with observation if $m<3$ \cite{ols87}.
The power spectrum of matter density perturbations does not appear
to be greatly modified by a decaying $\Lambda$-term,
at least for $0 \leq m \leq 2$ \cite{sil94}.
Lensing statistics favor models with $m \geq 1$ \cite{tor96},
or $m \geq 1.6$ when combined with other tests involving CMB
anisotropies and the classical magnitude-redshift relation for
high-redshift supernovae \cite{sil97}.
Other aspects of models in which $\Lambda$ decays as $a^{-m}$
have been discussed by several authors \cite{mai94,sis91,mat95},
although no specific numerical bounds are set on $m$.

The question of the initial singularity has so far received
little attention in theories of this kind \cite{ove98}.
Some explicitly nonsingular solutions have been constructed,
all with $m=2$ \cite{oze86}.  In one other case it is
noted in passing that the existence of an initial singularity would
require $0<m<4$ under some circumstances \cite{mat95}.
The remaining authors either do not mention the issue, or
(as in one case \cite{ols87}) rule out {\em a priori\/} the
possibility of nonsingular solutions.
In this paper, we take a broader view and examine all possible
solutions for the scale factor, including those in which it
takes a nonzero minimum value.
Moreover we will extend the discussion, not only to general $m$, 
but to general $\gamma$ as well [where $\gamma$ characterizes the
equation of state of ordinary matter, eq.~(\ref{eqState})].
If $m$ and $\gamma$ are thought of as defining a parameter space, 
then we determine, firstly, the extent to which the space is
singularity-free; and secondly, the extent to which it is 
observationally viable.

\subsection{Evolution of the Scale Factor}
\label{sec:evol}

We begin with the dynamical equations~(\ref{LemEq}) -- (\ref{mainDE}).
In particular, we consider the energy conservation law~(\ref{genECE}), 
which, with the decay law~(\ref{Bam}), becomes:
\begin{equation}
\frac{d}{da} \left( \rho a^{3\gamma} \right) = \left( \frac{m{\cal B}}{8\pi G}
                                               \right) a^{3\gamma-(m+1)} .
\end{equation}
Integrating, we find for the matter energy density:
\begin{equation}
\rho (a) = \rho_0 a^{-3\gamma} f(a) ,
\label{NewRho}
\end{equation}
where we have set $a(t_0)=a_0=1,\rho(a_0)=\rho_0$, and defined:
\begin{eqnarray}
f(a) & \equiv & 1 + \kappa_0 \times \left\{ \begin{array}{ll}
                    \frac{ {\textstyle m(a^{3\gamma-m}-1)} }
                         { {\textstyle 3\gamma-m} }
                          & \mbox{ if } m \neq 3\gamma \\
                    3\gamma \ln(a) 
                          & \mbox{ if } m=3\gamma
                    \end{array} \right. 
                    \label{fDefn} \\
\kappa_0 & \equiv & {\cal B}/8\pi G \rho_0 .
                    \label{kappaDef1}
\end{eqnarray}
When $m=0$, then $f(a)=1$ and eq.~(\ref{NewRho}) yields the usual result
that $\rho$ scales as $a^{-3}$ in a pressure-free universe ($\gamma=1$)
and $a^{-4}$ in a radiation-dominated universe ($\gamma=4/3$).
The new parameter $\kappa_0$ can be fixed in terms of observable 
quantities by means of the decay law~(\ref{Bam}), which gives
${\cal B} = \Lambda_0 = 3H_0^2\lambda_0$
[with $a_0=1$ and $\lambda_0$ defined as usual by eq.~(\ref{lamOm0defns})].
Substituting this result into eq.~(\ref{kappaDef1}), we find:
\begin{equation}
\kappa_0 = \lambda_0/\Omega_0 ,
\label{kappaDefn}
\end{equation}
where $\Omega_0$ is defined by eq.~(\ref{lamOm0defns}).
The parameter $\kappa_0$ is simply the ratio of energy density in the
cosmological term to that in ordinary matter at the present epoch.

Substitution of eqs.~(\ref{Bam}) and (\ref{NewRho})
into the Lema\^{\i}tre equation~(\ref{LemEq}) yields:
\begin{equation}
\frac{da}{d\tau} = a \left[ \Omega_0 a^{-3\gamma} f(a) -
                   (\Omega_0+\lambda_0-1) a^{-2} +
                   \lambda_0 a^{-m} \right]^{1/2} ,
\label{Expansion}
\end{equation}
where we have made use of the definitions~(\ref{lamOm0defns}),
recalled that $\dot{a}/a = (H_0/a) \, da/d\tau$, and selected the 
positive root since redshifts rather than blueshifts are observed.

At this point we could choose integer values of $m$ and
attempt to solve analytically for $a(\tau)$, as in \S\ref{sec:t}.
Detailed analyses have been carried out along these lines for the case
$m=0$; ie., for a constant cosmological term \cite{Exact}.
It is doubtful that they can be usefully extended 
to the general situation in which $m \neq 0$.
We opt instead to solve the problem numerically, 
following the lead of Felten and Isaacman \cite{fel86}.
The time-derivative of eq.~(\ref{Expansion}) is:
\begin{equation}
\frac{d^{\, 2}a}{d\tau^2} = \left( 1 - \frac{3\gamma}{2} \right)
                            \Omega_0 a^{1-3\gamma} f(a) +
                            \lambda_0 a^{1-m} .
\label{Deceleration}
\end{equation}
[This could equally well have been obtained from eq.~(\ref{accel}).]
We substitute eqs.~(\ref{Expansion}) and (\ref{Deceleration}) into 
a Taylor expansion for the scale factor:
\begin{equation}
a_k \, \approx \, a_{k-1} + \left( \frac{da}{d\tau} \right)_{k-1} 
               \! \! \! \! \! \Delta\tau + \frac{1}{2} 
               \left( \frac{d^{\, 2}a}{d\tau^2} \right)_{k-1} 
               \! \! \! \! \! (\Delta\tau)^2 .
\label{TaylorExp}
\end{equation}
This can be integrated numerically backwards in time to determine
whether or not a model with given values of $\{m,\gamma,\Omega_0,\lambda_0\}$ 
eventually reaches $a=0$.
We have tested the procedure for the case of a constant cosmological term
($m=0$) and dust-like equation of state ($\gamma=1$), and our results
in this case confirm those of Felten and Isaacman \cite{fel86}.
Fig.~\ref{figure5} depicts a group of examples with $\Omega_0=0.34$
(typical of large-scale observations \cite{bah97}).
and various values of $\lambda_0$ (labelled beside each curve).  
Note that the difference between this figure and the ones in the previous
section is that we now include models of all {\em three\/} kinds:  closed
(long-dashed lines), flat (dash-dotted lines) and open (short dashes).
(To keep the diagram from being too crowded, we show only models with
the same value $\Omega_0$ of the matter density.)
Fig.~\ref{figure5} indicates that negative values of $\lambda_0$ can lead to 
recollapse in open, as well as flat universes ({\em cf.\/} \S\ref{sec:t}).
Of greater interest, however, is the fact that
models with $\lambda_0$ above a {\em critical value\/} $\lambda_{\ast}$
(=1.774 605 in this case) avoid the big bang, undergoing a
finite ``big bounce'' instead.  Models with slightly less than 
this critical value (eg., $\lambda_0=1.76$ in this case) are of the 
``coasting Lema\^{\i}tre'' kind:
they begin in a singular state but go through an extended phase 
in which the scale factor is nearly constant.  
Models with {\em exactly\/} $\lambda_0=\lambda_{\ast}$
(shown in Fig.~\ref{figure5} with a solid line) are perhaps
the most interesting of all.  As time $\tau \rightarrow -\infty$,
they neither plunge to zero size nor bounce back up to infinite size,
but level off indefinitely at a constant value $a=a_{\ast}$ 
(=0.46 in this case).  These are nonsingular Eddington-Lema\^{\i}tre 
models, asymptotic to the static Einstein universe in the infinite past.

All these features of the $m=0,\gamma=1$ models have been discussed at
greater length elsewhere \cite{fel86}.  Our purpose here is to generalize
the discussion to arbitrary values of $m$ and $\gamma$.

\subsection{Critical Values of {\boldmath $\lambda_0$}}
\label{sec:critical}

In particular, we wish to obtain general expressions for 
the critical value $\lambda_{\ast}$ of the lambda parameter
$\lambda_0$ and minimum size $a_{\ast}$ of the scale 
factor, given any class of models $\{m,\gamma,\Omega_0\}$.
As discussed in \S\ref{sec:time}, 
we are interested in models for which $da/d\tau \rightarrow 0$ at
some time in the past.  This occurs, for example, at the moment of 
the ``bounce'' in all the oscillating models shown in Fig.~\ref{figure5}.
The {\em critical\/} case is distinguished by that fact that not only
$da/d\tau$, but also $d^{\, 2}a/d\tau^2$ vanishes at this
point \cite{fel86,bor88}.
We therefore set $da/d\tau=d{\, ^2}a/d\tau^2=0$ in 
eqs.~(\ref{Expansion}) and (\ref{Deceleration}).  This yields:
\begin{equation}
\lambda_{\ast} = \frac{(3\gamma-2)(3\gamma-m)\Omega_0}
                 {3\gamma(2-m)a_{\ast}^{3\gamma-m}+(3\gamma-2)m} \, ,
\label{CritVal}
\end{equation}
where $a_{\ast}$, the minimum value of the scale factor, 
is found by solving:
\begin{eqnarray}
3\gamma (3\gamma-m)\Omega_0 a_{\ast}^2 + 3\gamma (2-m)(1-\Omega_0) 
                         a_{\ast}^{3\gamma} & - & \nonumber \\
- (3\gamma-2)(3\gamma\Omega_0-m) a_{\ast}^m & = & 0 \, .
\label{Cubic}
\end{eqnarray}
In general eq.~(\ref{Cubic}) has to be solved numerically, but in the case
$m=0, \gamma=1$ it reduces to a cubic equation \cite{fel86}.
Fig.~\ref{figure6} is a phase space portrait of this case, with each point
on the diagram corresponding to a choice of $\Omega_0$ and $\lambda_0$ 
(after Lahav {\em et al\/} \cite{lah91}).
The critical values $\lambda_{\ast}$ are represented in this figure by a 
heavy solid line.
The region to the right of this curve corresponds to universes with 
$\lambda_0 > \lambda_{\ast}$; that is, with no big bang.
Also shown in Fig.~\ref{figure6} is a straight dash-dotted line 
representing the boundary between open and closed universes; 
models on this line have $k=0$ ($\Omega_0=1-\lambda_0$) while 
those on the left and right have $k<0$ and $k>0$ respectively 
(\S\ref{sec:defns}).

We now have the tools we need to investigate models with arbitrary
values of $m$ and $\gamma$.  The idea will be to use phase space diagrams
like Fig.~\ref{figure6} to determine how much of the parameter space
(1) corresponds to models with a nonzero minimum scale factor; and
(2) agrees with observational constraints.
We can then confirm whether a given model with $\{m,\gamma,\Omega_0\}$
does in fact avoid the big bang by carrying out the numerical integration
described in \S\ref{sec:evol}, and plotting the results on 
evolution diagrams like Fig.~\ref{figure5}.  

\subsection{Observational Constraints}

\subsubsection{Upper Bounds on $\lambda_0$}
\label{sec:upperBounds}

We pause first to take stock of some of the experimental constraints
that have been placed on models with nonzero cosmological terms.
Most immediate are direct upper bounds on $\lambda_0$ from a variety
of methods, most of them assuming that $\Omega_0+\lambda_0=1$.
Until recently, these have typically been of order $\sim 1$ \cite{lah91}.
Additional methods, however, have become available in the past few years.
CMB fluctuations, for instance, have 
produced an upper limit of $\lambda_0 \leq 0.86$ \cite{bun95}.
Gravitational lens statistics give a tighter bound of
$\lambda_0<0.66$ \cite{koc96},
and observations of Type Ia supernovae appear to reduce this still 
further, to $\lambda_0<0.51$ \cite{per97}.
All of the above are described as 95\% confidence level measurements.
On the other hand, a {\em lower\/} limit of $\lambda_0>0.53$ has been obtained
from the galactic luminosity density --- {\em also\/} at 95\% confidence 
\cite{tot97}.

Complicating the picture somewhat is the fact that several other
observational data are well-explained by substantial values of $\lambda_0$,
including the lack of observed small-scale dark matter,
the expectation that inflation should lead to near-flatness,
and especially the high {\em age\/} of the Universe inferred from
models of stellar evolution.
This ``cosmic concordance'' \cite{koc96} at one time 
led to calls for $\lambda_0 \sim 0.8$ \cite{efs90}, although this
has since dropped to $0.5-0.7$ \cite{kra95}.

Thus, the observational situation is not yet settled.
It is clear, however, that the nonsingular models in Fig.~\ref{figure5},
which require $\lambda_0 > 1.77$, are almost certainly unphysical.
It remains to be seen if the same conclusion applies when
$m \neq 0$ and/or $\gamma \neq 1$.

\subsubsection{Age of the Universe}
\label{sec:Ages}

A lower limit on $\lambda_0$ derives from the age of the Universe,
$t_0 = \int_0^{a_0} da/\dot{a}$.  In our units of Hubble times:
\begin{equation}
\tau_0 \equiv H_0 t_0 = \int_0^1 \frac{da}{da/d\tau} .
\end{equation}
If we use a recent value of $73 \pm 10$~km~s$^{-1}$~Mpc$^{-1}$ for $H_0$ 
\cite{fre96} in conjunction with the oldest globular cluster age of 
$t_0 > 9.6$~Gyr \cite{cha97}, then $\tau_0 \geq 0.62$;
that is, the Universe is at least 0.62 Hubble times old.
Of course, as noted in \S\ref{sec:Casel=1}, the true status of
these two parameters is still a subject of some controversy.
A lower value of $H_0 = 55 \pm 10$~km~s$^{-1}$~Mpc$^{-1}$ \cite{san96}
leads to $\tau_0 \geq 0.44$.
We choose an the intermediate value of 0.5.
In conjunction with our expression~(\ref{Expansion})
for $da/d\tau$, this implies:
\begin{equation}
\tau_0 = \int_1^\infty \frac{dv}{ \left[ \Omega_0 v^{2+3\gamma} g(v) -
                     ( \Omega_0+\lambda_0-1 ) v^4 +
                       \lambda_0 a^{2+m} \right]^{1/2} } \; > \; 0.5 ,
\label{Age-Int}
\end{equation}
where we have changed variables of integration from $a$ to
$v \equiv a^{-1} = 1+z$ for computational reasons, and:
\begin{equation}
g(v) \equiv 1 + \kappa_0 \times \left\{ \begin{array}{ll}
                \frac{ {\textstyle m(v^{m-3\gamma}-1)} }
                     { {\textstyle 3\gamma-m} }
                     & \mbox{ if } m \neq 3\gamma \\
               -3\gamma \ln(v)
                     & \mbox{ if } m=3\gamma .
                \end{array} \right.
\end{equation}
Eq.~(\ref{Age-Int}) reduces to the standard results \cite{ref67,lah91}
in the case $m=0,\gamma=1$.  

Numerical solution of eq.~(\ref{Age-Int}) produces a lower limit 
on $\lambda_0$ as a function of $\{m,\gamma,\Omega_0\}$.
This age constraint is shown in Fig.~\ref{figure6} 
(for the case $m=0, \gamma=1$) as a long-dashed line.
Its position matches that in a similar plot by
Lahav {\em et al\/} \cite{lah91}.
The region to the left of this curve corresponds to universes 
younger than half a Hubble time.
Big bangless models, of course, are not constrained by this;
they are infinitely old (by definition), the heavy solid line
being precisely the boundary where $\tau_0 \rightarrow \infty$.
The main impact of the age constraint is to rule out models
with a {\em negative\/} cosmological term.

In Fig.~\ref{figure7}, we show the effects of
varying the parameters $m$ and $\gamma$ respectively on this age
constraint.
It may be seen that altering the value of $m$ changes the
slope of the curve, but does not otherwise greatly
affect the age, even over the range $-1 \leq m \leq 3$.
Altering the value of $\gamma$, on the other hand, has a larger
effect.  In particular, the ``harder'' the equation of state
(ie., the larger the value of $\gamma$), the further this 
constraint encroaches on the available parameter space.
This is in accord with the well-known fact that a radiation-dominated
universe ($\gamma=4/3$), for example, is a short-lived one.

\subsubsection{Gravitational Lensing and the Antipode}

For closed models, the most stringent constraint on $\lambda_0$ comes from
gravitational lensing, which requires that the ``antipode'' 
be further away than the most distant normally lensed object \cite{got89}.
The antipode is the point where the radial coordinate 
$\omega \equiv \int_t^{t_0} dt/a(t) = \int_a^{a_0} da/\dot{a}a$
takes the value $\pi$ \cite{mcv65}.
Using our expression~(\ref{Expansion}) for $da/d\tau$, 
together with eq.~(\ref{Useful}), we obtain (taking $a_0=1$ as usual):
\begin{equation}
\omega = \sqrt{\Omega_0+\lambda_0-1} \; \int_1^{1+z} \frac{dv}
         { \left[ \Omega_0 v^{3\gamma} g(v) - ( \Omega_0+\lambda_0-1 ) v^2 +
         \lambda_0 a^{m} \right]^{1/2} } ,
\label{Lens-Int}
\end{equation}
where we have again changed integration variables from $a$ to $v$.
The standard formula \cite{ref67} is recovered when $m=0$ and $\gamma=1$.

At present, the furthest known normally lensed object
is a pair of lensed galaxies at $z_{\ell}=4.92$ \cite{fra97}.
We therefore require:
\begin{equation}
z_A(\Omega_0,\lambda_0,m,\gamma) > 4.92 ,
\end{equation}
where $z_A(\Omega_0,\lambda_0,m,\gamma)$ is defined by eq.~(\ref{Lens-Int})
with $\omega=\pi$.  Numerical solution of this equation yields
an upper limit on $\lambda_0$ as a function of $\{m,\gamma,\Omega_0\}$.
This lensing constraint is shown in Fig.~\ref{figure6} 
(for the case $m=0, \gamma=1$) as a short-dashed line.
Its position is close to that in previous plots \cite{lah91,got89}
employing a smaller value $z_{\ell}=3.27$ (our constraints are slightly 
stronger).  The region to the right of this curve corresponds to universes 
incompatible with the lensing observations.
Since this includes the entirety of nonsingular parameter space, we can
see that oscillating models with constant $\Lambda$ and zero pressure
are ruled out, as noted previously by these authors.

In Fig.~\ref{figure8}, we show the effects of varying the parameters 
$m$ and $\gamma$ respectively on this lensing constraint.
As before, it is seen that harder values of $\gamma$ lead to tighter
constraints on the available parameter space.  However, the situation
with regard to $m$ is altered quite dramatically.  In particular, the
higher the value of $m$, the {\em weaker\/} the lensing constraint becomes.
As we will see, this significantly improves the prospects for viable
big bangless models.

\subsubsection{The Maximum Redshift Constraint}

A fourth observational constraint, which must be satisfied {\em only\/}
by nonsingular models, concerns the maximum observable redshift
$z_{\ast} = a_{\ast}^{-1}-1$ in a universe with a minimum 
scale factor $a_{\ast}$.
This must obviously be larger than the greatest redshift $z_{obs}$ actually 
observed.  Thus the nonsingular models in Fig.~\ref{figure5},
which never get smaller than $a_{\ast}=0.46$, cannot accommodate
redshifts greater than $z_{\ast}=1.2$.  This disagrees with observations
such as those of the lensed galaxies mentioned above.  

The condition $z_{\ast} > z_{obs}$ can be reformulated as an upper limit on 
the {\em matter density\/} of the Universe.  Assuming that
$da/d\tau=0$ and $d\,^2a/d\tau^2 \geq 0$ at $z=z_{\ast}$ for nonsingular 
models, eqs.~(\ref{Expansion}) and (\ref{Deceleration}) can be combined
to read (for $m \neq 3\gamma$):
\begin{equation}
\Omega_0 \leq \frac{2-m+(m/3\gamma)(3\gamma-2)(1+z_{\ast})^{3\gamma-m}}
                   {2-m-(3\gamma-m)(1+z_{\ast})^{3\gamma-2}
                       +(3\gamma-2)(1+z_{\ast})^{3\gamma-m}} .
\label{MaxRedshift}
\end{equation}
Eq.~(\ref{MaxRedshift}) reduces to earlier expressions of 
B\"orner and Ehlers \cite{bor88} when $m=0$ and $\gamma=1$ or $4/3$.
These authors have then argued as follows:
given that quasar redshifts have been observed out to $z_{obs}>4$, 
we know that the Universe has $z_{\ast}>4$.  This constraint with 
eq.~(\ref{MaxRedshift}) implies 
(assuming $m=0$ and $\gamma=1$ or $4/3$) that 
$\Omega_0 \leq 0.018$, which is contrary to observation.
This indicates that our Universe could not have been nonsingular.

Let us see how the above conclusion changes when we generalize the situation
to values of $m \neq 0$.  Moreover, we will {\em strengthen\/} 
the argument by noting that some distant galaxies have
now been assigned photometric redshifts as high as $z_{obs}>6$
\cite{lan96}, implying that $z_{\ast}>6$.
The resulting upper limits on density $\Omega_0$ are listed in 
Table~\ref{table2} for various values of $m$ and $\gamma$.
From this table we see that the new photometric redshifts tighten the
B\"orner-Ehlers constraint noticeably:  as long as $m=0$, a nonsingular
Universe requires $\Omega_0 \leq 0.006$ (if $\gamma=1$) or $\Omega_0 < 0.001$
(if $\gamma=4/3$).  These numbers, of course, are too low to describe the 
real universe.  However, Table~\ref{table2} demonstrates that much higher
densities are possible in singularity-free models with {\em variable\/} 
$\Lambda$.
For example, retaining pressure-free conditions ($\gamma=1$), we see
that if $m=1$, then the matter density in a nonsingular universe must 
satisfy $\Omega_0 < 0.48$.  This value is not unreasonable at all;
in fact it is well above most dynamical measurements, which
suggest $\Omega_0 \approx 0.3$ \cite{bah97}.
The constraint is similarly loosened if we move toward softer
equations of state with $\gamma <1$, such as those that have been
proposed in \cite{Gamma}.

\subsection{Viable Oscillating Models}
\label{Whyk=+1}

We now demonstrate that models with $m \neq 0$ are capable of satisfying
all the constraints discussed above.
Fig.~\ref{figure9} is a phase space diagram like Fig.~\ref{figure6}, 
but plotted for a range of nonzero values of $m$.
We have assumed $\gamma=1$ as before, but there is now a different line of 
critical values $\lambda_{\ast}(m,\gamma,\Omega_0)$ for each value of $m$.
Like Fig.~\ref{figure6}, Fig.~\ref{figure9} shows that much of the nonsingular 
parameter space (below and to the right of the heavy solid lines) is eliminated 
because it does not overlap with the regions allowed by the lensing constraint
(above and to the left of the lighter dashed lines).
This is especially true for small values of $m$.
{\em With increasing\/} $m$, however, significant triangle-shaped regions 
of parameter space appear near $k=0$.
Thus for $m=1$, there are allowed models with $\Omega_0 \sim 0.3$ and 
$\lambda_0$ between about 0.7 and 0.9.
These are very close to the values favored by observation 
\cite{bah97,kra95}, so we focus on this case.  
The lensing constraint imposes the lower bound $\Omega_0 > 0.31$.
If we take $\Omega_0=0.34$ as a specific example (as before),
then by tracing horizontally across Fig.~\ref{figure9} we find that lensing
also places an upper bound $\lambda_0 \leq 0.72$ on the cosmological term.
From eq.~(\ref{CritVal}) the critical value for this case turns out to be
$\lambda_{\ast}=0.68$, which is marginally consistent with the upper
limits on $\lambda_0$ mentioned earlier.  Therefore $m=1$ models 
with $\Omega_0=0.34$ and $0.68 \leq \lambda_0 \leq 0.72$ are both realistic
and singularity-free.  

For larger values of $m$, the range of acceptable
$\lambda_0$-values is broader, but one is also driven to higher values of
$\Omega_0$.  With $m=3/2$, for example, we find that viable nonsingular models 
occur only for $\Omega_0 > 0.46$, and that at $\Omega_0=0.51$ they lie in 
the range $0.51 \leq \lambda_0 \leq 0.57$.
At $m=7/4$ and $\Omega_0=0.59$ this broadens to 
$0.42 \leq \lambda_0 \leq 0.50$ (low enough to satisfy 
even the supernova constraint \cite{per97}).
As $m \rightarrow 2$, we find that $a_{\ast} \rightarrow 0$,
so that nonsingularity is lost.  This also happens in the limit
$\gamma \rightarrow 2/3$ and, interestingly, as 
$\lambda_{\ast} \rightarrow 1-\Omega_0$.  (It is for this latter reason 
that all the nonsingular solutions in Fig.~\ref{figure9} have $k>0$.)

Unusual values of $\gamma$, although interesting and possibly relevant 
at early times, do not alter these results in any fundamental way.
In practice we find that the parameter space allowed by observation 
shrinks slightly for harder values of $\gamma$, like $4/3$, 
but grows significantly for softer values such as those in the range 
$0.4 < \gamma < 1$ \cite{Gamma}.

To verify that models with the properties described above can in fact
avoid the big bang, Fig.~\ref{figure10} shows the evolution of the
scale factor when $m=1, \gamma=1$ and $\Omega_0=0.34$, obtained by
numerical integration using the Taylor expansion~(\ref{TaylorExp}),
as before.  Various values of
$\lambda_0$ are labelled beside the appropriate curves.
This plot has exactly the same format as Fig.~\ref{figure5}, with long
dashes corresponding to $k>0$, short ones to $k<0$, and the dash-dotted
line corresponding to $k=0$. 

Fig.~\ref{figure10} confirms that when $\lambda_0$ takes on the critical value 
$\lambda_{\ast}$ (=0.679 807 621 in this case), the scale factor $a$ evolves 
back to a constant value $a_{\ast}$, as expected (solid line).
More importantly, the small {\em size\/} of this minimum value, 
$a_{\ast}=0.0097$, means that we can now accommodate 
observed redshifts up to $z_{\ast}=102$,
well beyond the furthest objects yet seen.  The model is thus compatible
with all observational data.  It cannot, however, accommodate larger redshifts 
like that attributed to the last scattering surface ($z_{lss} \sim 1100$),
let alone the era of nucleosynthesis, which occurred at temperatures 
$T_{nuc} \sim 10^{10}$~K, or --- since $T \propto (1+z)$ --- redshifts 
$z_{nuc} \sim 10^{10}$.
In fact, with $z_{\ast}=102$ the $m=1$ model heats up to no more than about
$T_{\ast} = 103 T_0 = 281$~K, which is just under room temperature.
That is to say, if the Universe is described by a nonsingular
variable-$\Lambda$ theory with $m=1$, then the CMB and the abundance 
of the light elements cannot be accounted for in the conventional way.

Unconventional explanations have been proposed for both these phenomena,
but are not widely accepted \cite{pee93}.
A more conservative approach might be to retain the traditional hot, dense
early phase in the context of variable-$\Lambda$ theory by moving to
larger values of $m$.
As discussed above, this leads to smaller values of $a_{\ast}$ 
and hence larger values of $z_{\ast}$ and $T_{\ast}$.
In this picture the high temperature of the early universe is a result,
not of an initial big bang singularity, but of a very deep ``big bounce.''
With $m=1.5$ and $\Omega_0=0.51$, for instance, we find that
$a_{\ast}=1.7 \times 10^{-4}$, corresponding to a maximum redshift
of 5900 and a ``bounce temperature'' of 16,000~K --- 
more than sufficient for recombination.  At $m=1.8$, this latter
number climbs to $5.7 \times 10^9$~K --- hot enough for 
nucleosynthesis.  (An analysis like that below Fig.~\ref{figure9}
shows that models with $m=1.8$ and $\Omega_0=0.61$, for example,
are observationally viable if $0.41 \leq \lambda_0 \leq 0.49$.)
With $m=1.9$, one can entertain bounce temperatures as high
as $2.9 \times 10^{13}$~K (2.5~GeV), approaching 
the realm of electroweak unification.
(Models with $m=1.9$ and $\Omega_0=0.67$ agree with observation if 
$0.39 \leq \lambda_0 \leq 0.48$.)
As $m\rightarrow 2$, in fact, one finds that $a_{\ast} \rightarrow 0$
and $T_{\ast} \rightarrow \infty$.
As remarked above, however, these higher values of $m$ come
at the modest observational price of higher matter densities.
The lensing constraint sets a limit of $\Omega_0 > 0.54$ for $m=1.8$ models, 
for example, and $\Omega_0 > 0.57$ for $m=1.9$ models.

\section{{\boldmath $\Lambda$} as a Function of the Hubble Parameter}
\label{sec:H}

\subsection{Previous Work}

We have seen in \S\ref{sec:a} that variable-$\Lambda$ theories
in which $\Lambda \propto a^{-m}$ appear to offer the possibility of
avoiding the big bang singularity without violating any observational 
constraints.  The parameter space occupied by viable models, however,
remains small.  As discussed in \S\ref{Whyk=+1}, one reason for this is
that the oscillating solutions are all closed,
whereas observational evidence tends to favor an open universe \cite{pee93}.
In this section we shift our attention to decay laws of the
form set out in eq.~(\ref{H-Law}):
\begin{equation}
\Lambda = {\cal C} \, H^{\, n} .
\label{CHn}
\end{equation}
Among other things we will find that a cosmological term of this kind
allows for a nonzero minimum scale factor in {\em open\/} models.

Terms of the form~(\ref{CHn}) are nearly as common as those studied 
in \S\ref{sec:a}, appearing in almost a third of the decay laws listed in 
Table~\ref{table1}.
The value $n=2$, favored on dimensional \cite{car92}
and other grounds \cite{lim94a}, is adopted almost universally
\cite{car92,arb94,wag93,sal93,lim94a,wet95,arb97}.
Other powers of $H$ are considered in three cases, but only in
combinations such as $\exp(-\int a H^n \, dt)$ \cite{reu87},
$a^{-2}H$ \cite{lim94b}, and $aH(dH/da)$ \cite{nes97}.
The question of the initial singularity has received little
attention in these papers, apart from one explicitly nonsingular
\cite{arb94} and one asymptotically de~Sitter-like
solution \cite{lim94b}.

\subsection{Riccati's Equation}
\label{sec:RiccatiAgain}

Substituting the decay law~(\ref{CHn}) into the differential
equation~(\ref{mainDE}), we obtain:
\begin{equation}
\frac{dH}{da} = \left( \frac{\gamma {\cal C}}{2a} \right) H^{\, n-1} -
                \left( \frac{3\gamma}{2a} \right) H -
                \left( \frac{3\gamma-2}{2a} \right)\left( \frac{k}{a^2} 
                \right) H^{-1} ,
\label{LooksLikeRiccati}
\end{equation}
where we have used the fact that $\ddot{a}/a = aH(dH/da)+H^2$.
This is nonlinear, but bears some resemblance to the Riccati 
equation~(\ref{RiccatiDE}).
We switch dependent variables from $H$ to $x \equiv H^{\, r}$, 
where $r$ is an arbitrary constant whose value will be chosen in a moment.
In terms of $x$, eq.~(\ref{LooksLikeRiccati}) takes the form:
\begin{equation}
\frac{dx}{da} = \left( \frac{r\gamma {\cal C}}{2a} \right) x^{\ell} -
                \left( \frac{3r\gamma}{2a} \right) x -
                \left[ \frac{r(3\gamma-2)k}{2a^3} \right] x^m ,
       \label{AlmostThere}
\end{equation}
where $\ell \equiv (n+r-2)/r$ and $m \equiv (r-2)/r$.

When $n=4$, we see that the choice $r=2$ puts $\ell=2$ and $m=0$,
whereupon eq.~(\ref{AlmostThere}) takes the form:
\begin{equation}
\frac{dx}{da} = {\cal P}(a) \, x^2 + {\cal Q}(a) \, x + 
                {\cal R}(a) ,
\label{ThatsIt}
\end{equation}
with:
\begin{equation}
{\cal P}(a) \equiv \frac{\gamma {\cal C}}{a} \; \; \; , \; \; \;
{\cal Q}(a) \equiv -\frac{3\gamma}{a} \; \; \; , \; \; \;
{\cal R}(a) \equiv \frac{(2-3\gamma)k}{a^3} .
\label{PQR}
\end{equation}
This is Riccati's equation~(\ref{RiccatiDE}), as desired.

When $n=2$, on the other hand, the choice $r=2$ leads to $\ell=1$ and $m=0$, 
whereupon eq.~(\ref{AlmostThere}) reduces to:
\begin{equation}
\frac{dx}{da} + {\cal S}(a) \, x = {\cal T}(a) ,
\label{JustLinear}
\end{equation}
with:
\begin{equation}
{\cal S}(a) \equiv \frac{\gamma (3-{\cal C})}{a} \; \; \; , \; \; \;
{\cal T}(a) \equiv \frac{(2-3\gamma)k}{a^3} .
\end{equation}
This case is linear, and may be solved easily.

For other values of $n$, a check reveals that no apparent choice of $r$
will put eq.~(\ref{AlmostThere}) into a form which can be solved
analytically for $x$ (ie., for $H$).
We therefore concentrate on the cases $n=2$ and $n=4$ 
for the time being, leaving the others to future numerical analysis.
In both cases the Hubble parameter is given by $H=\sqrt{x}$
(since $r=2$).

\subsection{The Case {\boldmath $n=2$}}
\label{1stOLDE}

Multiplying through by a factor of
$\exp \left[ \int {\cal S}(a)da \right] = a^{\gamma (3-{\cal C})}$
puts eq.~(\ref{JustLinear}) into exact form, which may be integrated 
directly for $x(a)$ and hence $H(a)$:
\begin{equation}
H(a) = \left\{ \left[ \frac{(2-3\gamma)k}{\gamma (3-{\cal C})-2} 
       \right] a^{-2} + C_0 a^{\gamma ({\cal C}-3)} \right\}^{1/2} ,
\label{HaSoln-1}
\end{equation}
where $C_0$ is a constant of integration.
Using the boundary condition $H(a_0)=H(1)=H_0$ to eliminate $C_0$,
we obtain:
\begin{equation}
H(a) = H_0 \left[ \alpha \, a^{\gamma ({\cal C}-3)} + \beta a^{-2} 
           \right]^{1/2} ,
\label{HaSoln-2}
\end{equation}
where:
\begin{equation}
\alpha \equiv 1 - \beta \; \; \; , \; \; \;
\beta \equiv \left[ \frac{2-3\gamma}{\gamma(3-{\cal C})-2} \right] 
             \frac{k}{H_0^2} .
\label{alphaBetaDefns}
\end{equation}
The parameters $\alpha,\beta$ and ${\cal C}$ can all be fixed in terms of 
observable quantities at the present time $t=t_0$, as follows.
With $n=2$, the decay law~(\ref{CHn}) gives
${\cal C} = \Lambda_0/H_0^2 = 3\lambda_0$,
where we have used the definition~(\ref{lamOm0defns}).
Substituting this result into eqs.~(\ref{alphaBetaDefns})
and (\ref{HaSoln-2}), we find:
\begin{eqnarray}
\alpha = \left[ \frac{ (3\gamma/2-1) \Omega_0 - \lambda_0 }
                     {(3\gamma/2)(1-\lambda_0) - 1} \right]\
             \; \; \; , \; \; \; \beta & = & 1 - \alpha ,
   \label{alphaBetaExpns} \\
\frac{da}{d\tau} = \left[ \alpha \, a^{2-3\gamma (1-\lambda_0)} +
                                 \beta \right]^{1/2} & , &
   \label{DAD}
\end{eqnarray}
where we have used eq.~(\ref{Useful}) and set $a_0=1$ as usual.

\subsection{Evolution of the Scale Factor}
\label{sec:minVals}

For special values of $\gamma$ and $\lambda_0$, it may be possible
to solve eq.~(\ref{DAD}) in terms of elliptic (or simpler) integrals.
For general purposes, however, we take the same approach as in 
\S\ref{sec:a} and evolve the scale factor numerically in terms of 
its first and second derivatives.  The latter of these is given by:
\begin{equation}
\frac{d^{\, 2}a}{d\tau^2} = \alpha \left[ 1 - \frac{3\gamma}{2}
                            (1-\lambda_0) \right] a^{1-3\gamma (1-\lambda_0)} .
\label{DDT}
\end{equation}
Eqs.~(\ref{DAD}) and (\ref{DDT}) can be substituted into the Taylor 
expansion~(\ref{TaylorExp}) and integrated backward numerically to
determine the behavior of the scale factor.

We also wish to determine the conditions under which the Universe
evolves backward to a nonzero minimum scale factor, $a=a_{\ast}$.
Setting $da/d\tau=0$ (at $a=a_{\ast}$), we find from eq.~(\ref{DAD}) that:
\begin{equation}
a_{\ast}^{3\gamma (1-\lambda_0)-2} = -\alpha/\beta .
\end{equation}
It is convenient to distinguish two cases,
according to whether the exponent on $a_{\ast}$ is positive
or negative.  Using eqs.~(\ref{alphaBetaExpns}),  we obtain:
\begin{equation}
a_{\ast} = \left\{ \begin{array}{ll}
           \left[ \frac{ {\textstyle 2\lambda_0-(3\gamma-2) \Omega_0 }}
                       { {\textstyle (3\gamma-2)(1-\Omega_0-\lambda_0) }} 
                       \right]^{1/[3\gamma(1-\lambda_0)-2]} 
                  & \mbox{ if } \lambda_0 < \lambda_c \\
           \left[ \frac{ {\textstyle (3\gamma-2)(1-\Omega_0-\lambda_0) }}
                       { {\textstyle  2\lambda_0-(3\gamma-2) \Omega_0 }} 
                       \right]^{1/[2-3\gamma(1-\lambda_0)]}
                  & \mbox{ if } \lambda_0 > \lambda_c ,
           \end{array} \right.
     \label{aSTAR}
\end{equation}
where we have defined $\lambda_c \equiv 1 - 2/3\gamma$
(leaving the case $\lambda_0=\lambda_c$ aside for the time being).

From eq.~(\ref{aSTAR}) we draw a number of important conclusions:
(1) Spatially {\em flat\/} solutions ($\Omega_0+\lambda_0=1$)
have either $a_{\ast}=\infty$ (if $\lambda_0<\lambda_c$)
or $a_{\ast}=0$ (if $\lambda_0>\lambda_c$).
The former case is not interesting.  
The latter case is de~Sitter-like, with the initial singularity pushed back
into the infinite past.  We have encountered this kind of solution
before (\S\ref{sec:deSitter}).

(2) With the modest assumption that $\gamma>2/3$
(ie., normal, non-inflationary matter), then we notice that 
{\em closed\/} solutions (ie., $\lambda_0>1-\Omega_0$) must satisfy
$\lambda_0 < (3\gamma/2 - 1) \Omega_0$,
while {\em open\/} ones ($\lambda_0<1-\Omega_0$) obey
$\lambda_0 > ( 3\gamma/2 - 1) \Omega_0$.
These conclusions follow from requiring that the terms in 
square brackets be positive; ie., from the requirement that
$a_{\ast}$ be a real number.  (The exception in which the
exponent is an even integer $\wp$ occurs only for special values
of the lambda parameter, $\lambda_0=1-2/3\gamma-1/6\wp\gamma$,
and will not be considered further here.)

(3) Requiring that $0 < a_{\ast} < 1$, 
we learn that both the numerators and denominators in eq.~({\ref{aSTAR}) 
must be positive.  Comparing their relative absolute magnitudes, 
we distinguish two possibilities:
(a) if $\lambda_0 < \lambda_c$, then both the
numerator and denominator must be positive, since otherwise
$(3\gamma-2)(1-\Omega_0-\lambda_0) < 2\lambda_0-(3\gamma-2)\Omega_0$,
which reduces to $\lambda_0 > \lambda_c$, contrary to the hypothesis.
On the other hand,
(b) if $\lambda_0 > \lambda_c$, then we {\em also\/}
find that both numerator and denominator must be positive,
since otherwise
$(3\gamma-2)(1-\Omega_0-\lambda_0) > 2\lambda_0-(3\gamma-2)\Omega_0$,
which reduces to $\lambda_0 < \lambda_c$, again contrary to the
hypothesis.  Therefore both the numerators and denominators are 
positive in all cases.

(4) It follows from the conclusion~(3), in conjunction with
the assumption $\gamma>2/3$, that:
\begin{equation}
\lambda_0 < 1-\Omega_0 .
\label{Conc4}
\end{equation}
In other words, to realistically describe the present Universe,
models with a nonzero minimum scale factor must, in the present theory,
be {\em open\/}.  While some nonsingular open solutions have been found 
in theories employing scalar fields and higher-order curvature terms 
\cite{bek74}, we are not aware of precedents for this in theories 
based on the cosmological term.

(5) By combining eq.~(\ref{Conc4}) with the conclusion~(2), 
we infer that:
\begin{equation}
\lambda_0 > (3\gamma/2-1) \Omega_0 ,
\label{MINlambda0}
\end{equation}
which sets a lower limit on the size of the cosmological term.

(6) Eqs.~(\ref{Conc4}) and (\ref{MINlambda0}) together impose 
an upper limit:
\begin{equation}
\Omega_0 < 2/3\gamma ,
\label{UpperLim}
\end{equation}
on the matter density of the Universe.

\subsection{Minimum Values of the Scale Factor}

The information contained in eqs.~(\ref{aSTAR}) -- (\ref{UpperLim})
is summarized in Fig.~(\ref{figure11}), which is a phase space plot
like Fig.~\ref{figure9}, but enlarged to show only the range of 
interest, $0 \leq \Omega_0 \leq 1$ and $0 \leq \lambda_0 \leq 1$.
Models with $(\lambda_0,\Omega_0)$ are represented by points on this diagram,
as usual.  The critical values of $\lambda_0$ in this theory define the
upper edges of the triangular region at the base of the diagram; ie., the
region bounded by the curves $\Omega_0+\lambda_0<1$ (dash-dotted line)
and $\Omega_0+\lambda_0>(3\gamma/2)\Omega_0$ (dashed line).
All models between these curves are nonsingular, with real values of 
$a_{\ast}$ in the range $0 < a_{\ast} < 1$, as we have stipulated.
Fig.~\ref{figure11} is plotted for $\gamma=1$.

Using eq.~(\ref{aSTAR}), we have plotted contours of equal minimum 
scale factor $a_{\ast}$ in this region (heavy solid lines).
Any point along one of the contours corresponds to an oscillating
model with the labelled value of $a_{\ast}$.
Following the discussion in \S\ref{Whyk=+1}, we would like to find 
models in which $a_{\ast}$ is as small as possible, in order to obtain
the largest possible maximum observable redshift
$z_{obs} \leq a_{\ast}^{-1}-1$.
For instance, to be compatible with observations of quasars
($z_{obs} \approx 10$), $a_{\ast}$ must be less than about 0.1.
If we wish to explain the CMB as relic radiation from the last scattering
surface at $z_{lss} \approx 1100$, then we require a smaller minimum
scale factor, $a_{\ast}<0.001$.  And to meet the demand that the early 
Universe heat up to nucleosynthesis temperatures 
[$T_{nuc} \approx T_0 \, (a_0/a_{\ast}) \approx 10^{10}$~K],
our model must satisfy:
\begin{equation}
a_{\ast} < 10^{-9} \; \; \; \mbox{ (nucleosynthesis) } .
\label{NucLimit}
\end{equation}
Fig.~\ref{figure11} demonstrates that the present theory
can readily satisfy this constraint.
Any model lying along the curve labelled $10^{-9}$ will be capable, 
in principle, of reaching these temperatures near the ``big bounce.''
This confirms comments made by several authors \cite{Bounce}
that there is no reason in principle why the oscillations in a 
nonsingular model cannot be deep enough to account for all the 
evidence which is usually taken as proof that the Universe began 
in a singularity.  Moreover, the only observational constraint which 
seriously limited the variable-$\Lambda$ models in the last section --- 
the lensing constraint --- does not apply in this section because there
is no antipode in an open universe.

To meet the condition~(\ref{NucLimit}), 
models in this theory must lie close to the upper edge of the 
triangular region in Fig.~\ref{figure11}.  To an extent this is
``fine-tuning.''  However, it also allows us to make very definite 
predictions (as in \S\ref{sec:a}) about the values of $\lambda_0$ 
that would be required in a realistic oscillating model.
As an example, let us consider the observationally favored value of
$\Omega_0=0.3$ \cite{bah97}, and let us assume $\gamma=1$ as usual.
Tracing horizontally across the line defined by $\Omega_0=0.3$ 
in Fig.~\ref{figure11}, we can see that the nucleosynthesis
condition~(\ref{NucLimit}) is met by only {\em two\/} values of
the cosmological term: $\lambda_0 \approx 0.15$ and $\lambda_0 \approx 0.7$.
Of these, the larger value is only marginally viable, being very close to
the observational upper bounds described in \S\ref{sec:upperBounds}.
The smaller value, however, is perfectly acceptable from an observational
standpoint.

In general, the theory predicts that the most likely value of 
$\lambda_0$ is either {\em just below\/}:
\begin{equation}
1 - \Omega_0 ,
\end{equation}
or else {\em just above\/}:
\begin{equation}
(3\gamma/2-1) \, \Omega_0 ;
\end{equation}
which is to say, just above $\Omega_0/2$ in a dust-like universe
($\gamma=1$).
The former situation might be preferable to some on theoretical
grounds \cite{kra95}, while the latter is in better agreement with 
the increasingly stringent observational upper limits on 
$\lambda_0$ (\S\ref{sec:upperBounds}).

To confirm that models with these features really do avoid the big bang,
the solutions can be evolved backward in time as before, 
using eqs.~(\ref{DAD}) and (\ref{DDT}) with the 
Taylor series expansion~(\ref{TaylorExp}).
The results of this procedure are shown in Fig.~\ref{figure12} for the 
case $\Omega_0=0.3$ and $\gamma=1$.
This diagram is an enlarged version of the evolution plots in 
\S\ref{sec:a} (Figs.~\ref{figure5} and \ref{figure10}), 
showing only the past three Hubble times.
Values of $\lambda_0$ are marked beside the appropriate curves.  
Several features can be noted.  

Firstly, the initial singularity is avoided for any value of $\lambda_0$ 
between 0.15 and 0.7, as expected on the basis of the phase space 
diagram, Fig.~\ref{figure11}.
In the limiting case where $\lambda_0=0.7$ exactly, which is spatially
flat, we see that $a_{\ast}=0$ (de~Sitter-like behavior), as expected
based on the discussion following eq.~(\ref{aSTAR}).

Secondly, Fig.~\ref{figure12} confirms that the value of $a_{\ast}$ 
is smallest near the critical values of $\lambda_0$: 0.064 for the 
$\lambda_0=0.6$ case (just below 0.7), and 0.018 for the 
$\lambda_0=0.2$ case (just above 0.15).
These numbers have been chosen for illustrative purposes;
smaller values of $a_{\ast}$ (with consequently larger bounce temperatures)
are obtained by letting $\lambda_0$ approach the critical values more closely.

Thirdly, this evolution plot gives us some information about the ages
of the models; that is, the elapsed time since the big bang
(or the big bounce, as appropriate).  It may be seen that, 
within the range $0.15 < \lambda_0 < 0.7$, larger values
of $\lambda_0$ correspond to older universes, as usual:
1.10~Hubble times for $\lambda_0=0.2$, and 1.70~Hubble times for
$\lambda_0=0.6$.  Even if $H_0$ takes on its largest
currently acceptable value of $83$~km~s$^{-1}$~Mpc$^{-1}$ \cite{fre96},
the ages of the two models are 13.0 and 20.1~billion years respectively ---
well above the globular cluster limit of 9.6~Gyr \cite{cha97}.

\subsection{The Case {\boldmath $n=4$}}

We proceed to the other case of interest, $n=4$, which consists
of the Riccati equation~(\ref{ThatsIt}) for $x(a)$.
This can be solved using standard techniques \cite{DEs}.
Switching dependent variables from $x$ to $y$ via
$x = (-1/{\cal P}y) \, dy/da$, we obtain:
\begin{equation}
\frac{d^{\, 2}y}{da^2} + \zeta a^{-1} \, \frac{dy}{da} + 
       \eta a^{-4} \, y = 0 ,
\label{NontrivialDE}
\end{equation}
where:
\begin{equation}
\zeta \equiv 1+3\gamma \; \; \; , \; \; \;
\eta \equiv \gamma {\cal C} (2-3\gamma) k .
\label{AlphaBet0}
\end{equation}
The parameter $\eta$ can be connected to observation as follows.
With $n=4$, the decay law~(\ref{CHn}) gives
${\cal C} = \Lambda_0/H_0^4 = 3\lambda_0/H_0^2$.
Substituting this result into the second of eqs.~(\ref{AlphaBet0}), 
we find:
\begin{equation}
\eta \equiv 3\gamma (2-3\gamma) \lambda_0 (\Omega_0+\lambda_0-1) ,
\end{equation}
where we have used eq.~(\ref{Useful}) and set $a_0=1$ as usual.

Eq.~(\ref{NontrivialDE}) is linear as desired, but not straightforward
because of its variable coefficients.
We can recast it in normal form by changing independent variables 
from $a$ to $z \equiv \int e^{-\phi(a)} da$, where
$\phi(a) \equiv \int (\zeta a^{-1}) da = \zeta \ln a$.
This procedure leads to the following differential equation for $y(z)$:
\begin{equation}
z^{\mu} \, \frac{d^{\, 2}y}{dz^2} + \nu \, y = 0 ,
\label{normalForm}
\end{equation}
where $z=(1-\zeta)^{-1} \, a^{1-\zeta}$ and:
\begin{equation}
\mu \equiv \frac{2(2-\zeta)}{1-\zeta} \; \; \; , \; \; \;
\nu \equiv \eta (1-\zeta)^{-\mu} .
\label{AlphaBet1}
\end{equation}
Somewhat surprisingly, eq.~(\ref{normalForm}) has the same form as the 
differential equation~(\ref{mainDE-t}) governing the solutions of the
$\Lambda \propto \tau^{-\ell}$ models in \S\ref{sec:t}.
(The two are identical if we put $z \rightarrow \tau$, 
$y \rightarrow x$, $\mu \rightarrow \ell$, and
$\nu \rightarrow -\alpha$.)

So we could in principle bring over all the results of \S\ref{sec:t},
for the cases $\mu=1,2,3,4$ at least.
However, combining the definitions~(\ref{AlphaBet0}) and (\ref{AlphaBet1}),
we find that $\mu = (2/3\gamma)(3\gamma-1)$; or 
$\gamma = [3(1-\mu/2)]^{-1}$.
Therefore solutions obtained in this way would correspond to equations 
of state with $\gamma$-values of $2/3,\infty,-2/3$ and $-1/3$ respectively.
These do not describe realistic forms of matter, 
at least not in the present universe \cite{Gamma}.
Conversely, values of $\gamma$ that {\em are\/} reasonable 
(such as $\gamma=1$ or $4/3$) correspond to non-integral values 
of $\mu$ (such as $4/3$ and $3/2$ respectively).
It is doubtful that eq.~(\ref{normalForm}) can be solved analytically 
in these cases.  We therefore leave the possibility that $n=4$ for 
future numerical analysis.

\section{{\boldmath $\Lambda$} as a Function of the Deceleration Parameter}
\label{sec:q}

\subsection{Evolution of the Scale Factor}

We turn finally to the last of our phenomenological decay 
laws, eq.~(\ref{q-Law}), writing it in the form:
\begin{equation}
\Lambda = {\cal D} \, \left( \frac{\ddot{a}}{a} \right)^{\, r} .
\label{Daar}
\end{equation}
As far as we are aware, no such dependence has previously been considered 
for the cosmological term.  However, it is a natural extension of the 
other decay scenarios considered so far.  There is no fundamental
difference between the first and second derivatives of the scale factor
that would preclude the latter from acting as an independent variable
if the former is acceptable.

Substituting the decay law~(\ref{Daar}) into eq.~(\ref{mainDE}),
we find:
\begin{eqnarray}
{\cal D} \left( aH\frac{dH}{da}+H^2 \right)^r 
         & = & \left( 3 - \frac{2}{\gamma} \right) 
               \left( H^2 + \frac{k}{a^2} \right) + \nonumber \\
         & + & \frac{2}{\gamma} \left( aH\frac{dH}{da}+H^2 \right) .
\label{messyDE}
\end{eqnarray}
We adopt the value $r=1$ for the remainder of \S\ref{sec:q}, 
since we would like to solve for the Hubble parameter in analytic form 
in order to make use of of the Taylor expansion~(\ref{TaylorExp}).
Eq.~(\ref{messyDE}) then takes the form:
\begin{equation}
\frac{dH}{da} = \frac{\gamma}{a} \left( 
                \frac{3-{\cal D}}{\gamma{\cal D}-2} \right) H + 
                \frac{k}{a^3} \left( 
                \frac{3\gamma-2}{\gamma{\cal D}-2} \right) H^{-1} .
\label{simplerDE}
\end{equation}
As in \S\ref{sec:RiccatiAgain}, let us make a change of dependent variables
from $H$ to $x \equiv H^{\, s}$, where $s$ is an arbitrary constant.
Eq.~(\ref{simplerDE}) then takes the form:
\begin{equation}
\frac{dx}{da} = \frac{s\gamma}{a} \left(
                \frac{3-{\cal D}}{\gamma{\cal D}-2} \right) x + 
                \frac{s \, k}{a^3} \left( 
                \frac{3\gamma-2}{\gamma{\cal D}-2} \right) x^{(s-2)/s} .
\end{equation}
If we choose $s=2$, this is reduced to the linear form~(\ref{JustLinear}),
with:
\begin{equation}
{\cal S}(a) \equiv 2\gamma \left( \frac{{\cal D}-3}{\gamma {\cal D}-2}
                           \right) a^{-1} \; \; \; , \; \; \;
{\cal T}(a) \equiv 2k \left( \frac{3\gamma-2}{\gamma{\cal D}-2} 
                      \right) a^{-3} .
\end{equation}
Multiplying through by a factor of
$\exp [ \int {\cal S}(a)da ] = a^{2\gamma({\cal D}-3)/(\gamma {\cal D}-2)}$
and solving exactly as in \S\ref{1stOLDE}, we obtain for the 
Hubble parameter:
\begin{equation}
H(a) = \left[ C_0 a^{-2\gamma ({\cal D}-3)/(\gamma {\cal D}-2) } - ka^{-2} 
       \right]^{1/2} ,
\label{HaSoln-3}
\end{equation}
where $C_0$ is a constant of integration.
Imposing the boundary condition $H(a_0)=H(1)=H_0$, we find with the help
of eq.~(\ref{Useful}) that 
$C_0 = H_0^2 + k = H_0^2 ( \Omega_0 + \lambda_0 )$.
We can also fix ${\cal D}$ in terms of observable quantities.
With $r=1$, the decay law~(\ref{Daar}) gives
${\cal D} = \Lambda_0 (\ddot{a}/a)_{t=t_0}
          = -\Lambda_0/H_0^2 \, q_0 = -3\lambda_0/q_0$,
where $q_0$ is the present value of the deceleration parameter,
and we have used the definitions~(\ref{lamOm0defns}) and (\ref{qDefn}).
Substituting this result into eq.~(\ref{HaSoln-3}) along with
eq.~(\ref{Useful}), and recalling that $H=(H_0/a)da/d\tau$, we find:
\begin{equation}
\frac{da}{d\tau} = \left[ \alpha \, a^{\xi} + \beta \right]^{1/2} ,
\label{dadtau}
\end{equation}
where:
\begin{equation}
\alpha = \left( \Omega_0 + \lambda_0 \right) \; \; \; , \; \; \;
\beta = 1 - \alpha \; \; \; , \; \; \;
\xi \equiv \frac{(2-3\gamma) q_0}{q_0 + (3\gamma/2) \lambda_0} .
\label{xiDefn}
\end{equation}
This expression, together with its time derivative:
\begin{equation}
\frac{d^{\, 2}a}{d\tau^2} = \frac{\xi}{2} (\Omega_0+\lambda_0) a^{\xi-1} ,
\end{equation}
can be substituted into the Taylor expansion~(\ref{TaylorExp}).

\subsection{Minimum Values of the Scale Factor}

As in \S\ref{sec:minVals}, we require that oscillating
models satisfy $da/d\tau=0$ at $a=a_{\ast}$.
In conjunction with eq.~(\ref{dadtau}), this implies:
\begin{equation}
a_{\ast}^{\xi} = \left( \frac{\Omega_0 + \lambda_0 - 1}
                             {\Omega_0+\lambda_0} \right) .
\end{equation}
Let us write this out explicitly using the last of eqs.~(\ref{xiDefn}).
As with eq.~(\ref{aSTAR}), we will find it convenient to
distinguish two possible cases:
\begin{equation}
a_{\ast} = \left\{ \begin{array}{ll}
           \left( \frac{ {\textstyle \Omega_0 + \lambda_0 }}
                       { {\textstyle \Omega_0 + \lambda_0 - 1 }} \right)^
                       { \left[ \frac{ {\scriptscriptstyle 
                                        1+(3\gamma\lambda_0/2q_0) }}
                                     { {\scriptscriptstyle 
                                        3\gamma-2 }} \right] }
                  & \mbox{ if } q_0 > q_c \\
           \left( \frac{ {\textstyle \Omega_0 + \lambda_0 - 1 }}
                       { {\textstyle \Omega_0 + \lambda_0 }} \right)^
                       { \left[ \frac{ {\scriptscriptstyle 
                                       -(3\gamma\lambda_0/2q_0)-1 }}
                                     { {\scriptscriptstyle 
                                        3\gamma-2 }} \right] }
                  & \mbox{ if } q_0 < q_c ,
           \end{array} \right.
     \label{qSTAR}
\end{equation}
where we have defined $q_c \equiv -( 3\gamma/2) \lambda_0$
(leaving the case $q_0=q_c$ aside for the time being).

We can draw a number of useful conclusions from the form of
eq.~(\ref{qSTAR}).  Firstly, 
(1) that spatially {\em flat\/} solutions ($\Omega_0+\lambda_0=1$)
again have either $a_{\ast}=\infty$ (if $q_0>q_c$)
or $a_{\ast}=0$ (if $q_0<q_c$).
This is just as in the previous section (\S\ref{sec:minVals}).

(2) Secondly, requiring real values for $a_{\ast}$
(subject to the same proviso about even-numbered integer exponents
as in \S\ref{sec:minVals}), we can conclude that 
{\em closed\/} solutions (ie., $\lambda_0>1-\Omega_0$) must satisfy
$\lambda_0 > -\Omega_0$,
while {\em open\/} ones ($\lambda_0<1-\Omega_0$) obey
$\lambda_0 < -\Omega_0$.
It follows that, if $\lambda_0$ is a positive quantity, as
observations almost certainly indicate (\S\ref{sec:Ages}),
then models with a nonzero minimum scale factor must, 
in the present theory, be {\em closed\/}.
We will assume that both these conditions hold in the
remainder of \S\ref{sec:q}.

(3) Thirdly, requiring that $0 < a_{\ast} < 1$ as before, 
we learn that the deceleration parameter $q_0$ satisfies:
\begin{equation}
q_0 < q_c .
\end{equation}
This follows from the fact that $|\Omega_0+\lambda_0-1|$ cannot be 
greater than $|\Omega_0+\lambda_0|$ (assuming that $\Omega_0$ and
$\lambda_0$ are both positive).

(4) Finally, combining the conclusion~(3) with the 
definition of $q_c$, we infer that:
\begin{equation}
q_0 < -(3\gamma/2) \lambda_0 .
\end{equation}
Assuming as we are that the cosmological term is positive, this
implies that the deceleration parameter must be {\em negative\/}
for a universe filled with normal matter ($\gamma>2/3$).

Unfortunately, the deceleration parameter $q_0$ remains among the 
most poorly-constrained quantities in observational cosmology.
Nevertheless, it is fair to say that the majority opinion among
cosmologists holds that $q_0$ is probably positive \cite{car97}.
The most recent experimental determination, obtained from Type Ia
supernovae, leads to a value of $q_0=0.385 \pm 0.36$ \cite{tri97}.
Since oscillating models in the present theory not only have
$q_0 < 0$, but $k > 0$ as well, they are somewhat disfavored in 
comparison to those of \S\ref{sec:a} and \S\ref{sec:H};
and we judge that this is a reasonable place to halt our 
investigation for the time being.

\section{Conclusions}
\label{sec:con}

We have examined the evolution of the scale factor $a(t)$
in the presence of a variable cosmological term $\Lambda$,
and also extended existing treatments by adopting a fairly
general equation of state for ordinary matter.

A number of new exact solutions for $a(t)$ have been obtained
in cases where $\Lambda \propto t^{-\ell}$.
These models are singular, but can be significantly older than
those in which $\Lambda=$~constant.
For odd values of $\ell$, the cosmological term must be negative (or zero) 
if the scale factor is to be real-valued.
Our conclusions may not extend to cases in which $k \neq 0$; 
this will require more detailed numerical analysis.

For a cosmological term that scales as $\Lambda \propto a^{-m}$,
we have solved numerically for the scale factor as a function of time,
and found that there are closed models which are compatible with
observation and contain no big bang.
This is in sharp contrast to the situation where $\Lambda=$~constant,
for which experimental evidence firmly establishes the existence
of an initial singularity.  
(The variation effectively allows one to obtain a large $\Lambda$-term 
where it is most important --- near the ``big bounce'' --- without 
the price of a large cosmological constant at present times.)
This appears not to have been widely appreciated, probably because
variable cosmological terms have so far been studied almost exclusively
in the context of the cosmological ``constant'' problem.
We have obtained constraints from experimental upper limits on 
$\Lambda_0$, as well as requirements of sufficient age, normal
gravitational lensing at high redshifts, and others.
As specific numerical examples, oscillating models
with zero pressure, $\Omega_0=\{0.34,0.51,0.61,0.67\}$ and 
$\lambda_0$ in the ranges $\{0.68-0.72,0.51-0.57,0.41-0.49,0.39-0.48\}$
are observationally viable if $m=1,1.5,1.8$ or $1.9$ respectively.
If the bounce is to be deep enough to generate the temperatures 
required by conventional nucleosynthesis, then $m \geq 1.8$.

We have also solved numerically for the scale factor when
$\Lambda \propto H^n$.  In this case we have found {\em open\/} models
which can account for the observational data despite their lack of an 
initial singularity.
In particular, oscillating models with zero pressure and values of 
$\lambda_0$ either just above $\Omega_0/2$ or just below $1-\Omega_0$
are viable if $\Omega_0 < 2/3$ and $n=2$.
(We have investigated only the cases $n=2$ and $4$ in detail.)
If $\Omega_0 \approx 0.3$ and $\lambda_0 \approx 0.15$ (or $0.7$),
for example, then the most recent ``big bounce'' could have been 
deep enough to account for phenomena such as the cosmic microwave 
background radiation and light element synthesis in a model with
$n=2$.

For a cosmological term that depends on the {\em deceleration parameter\/}
via $\Lambda \propto q^r$, we have solved only the case $r=1$.
Closed oscillating models are possible, but require that $q$ be
negative at the present time, if the cosmological term is positive.

\acknowledgments

We are grateful to R.~H.~Brandenberger, L.~M.~de~Menezes, T.~Fukui,
F.~D.~A.~Hartwick, W.~Israel and C.~J.~Pritchet for their comments,
and to the National Science and Engineering Research Council of Canada 
for financial support.

\narrowtext
\newpage

\begin{figure}
\caption{Evolution of the scale factor for flat models with 
   $\Lambda\propto\tau^{-1}$ and $\gamma=1$.  Values of $\lambda_0$ 
   are labelled beside each curve, and $\Omega_0=1-\lambda_0$ in each case.}
\label{figure1}
\end{figure}

\begin{figure}
\caption{Evolution of the scale factor for flat models with
   $\Lambda\propto\tau^{-2}$ and $\gamma=1$.  Values of $\lambda_0$ 
   are labelled beside each curve, and $\Omega_0=1-\lambda_0$ in each case.}
\label{figure2}
\end{figure}

\begin{figure}
\caption{Evolution of the scale factor for flat models with
   $\Lambda\propto\tau^{-3}$ and $\gamma=1$.  Values of $\lambda_0$ 
   are labelled beside each curve, and $\Omega_0=1-\lambda_0$ in each case.}
\label{figure3}
\end{figure}

\begin{figure}
\caption{Evolution of the scale factor for flat models with
   $\Lambda\propto\tau^{-4}$ and $\gamma=1$.  Values of $\lambda_0$
   are labelled beside each curve, and $\Omega_0=1-\lambda_0$ in each case.}
\label{figure4}
\end{figure}

\begin{figure}
\caption{Evolution of the scale factor for models with $m=0, \gamma=1$,
   $\Omega_0=0.34$, and values of $\lambda_0$ labelled beside each curve
   (after Felten and Isaacman [5]).}
\label{figure5}
\end{figure}

\begin{figure}
\caption{Phase space diagram showing constraints on models with $m=0$ 
   and $\gamma=1$ (after Lahav {\em et al\/} [15]).}
\label{figure6}
\end{figure}

\begin{figure}
\caption{The age constraint $\tau_0 > 0.5$:
         (a) as a function of $m$, assuming $\gamma=1$; and 
         (b) as a function of $\gamma$, assuming $m=0$.}
\label{figure7}
\end{figure}

\begin{figure}
\caption{The lensing constraint $z_A > 4.92$:
         (a) as a function of $m$, assuming $\gamma=1$; and 
         (b) as a function of $\gamma$, assuming $m=0$.}
\label{figure8}
\end{figure}

\begin{figure}
\caption{Enlarged view of the phase space diagram, Fig.~\ref{figure6}, 
   now plotted for various values of $m$ between 0 and 2 (labelled 
   beside each pair of curves), assuming $\gamma=1$.}
\label{figure9}
\end{figure}

\begin{figure}
\caption{Evolution of the scale factor for universes with $m=1, \gamma=1$ 
   and $\Omega_0=0.34$.  Compare Fig.~\ref{figure5}.}
\label{figure10}
\end{figure}

\begin{figure}
\caption{Phase space diagram for the case $n=2$ with $\gamma=1$,
         showing contours of equal minimum size $a_{\ast}$.}
\label{figure11}
\end{figure}

\begin{figure}
\caption{Evolution of the scale factor for models with $n=2, \gamma=1$,
   $\Omega_0=0.3$, and values of $\lambda_0$ labelled beside each curve.}
\label{figure12}
\end{figure}

\newpage

\begin{table}
\caption{Examples of phenomenological $\Lambda$-decay laws.}
\label{table1}
\begin{tabular}{lr}
Decay Law\tablenotemark[1] & Reference \\
\tableline
$\Lambda\propto t^{-2}$ & \cite{end77,can77,ber86,lau85,ber91,bee94,lop96} \\
$\Lambda\propto T^{\, 4}$ & \cite{can77} \\
$\Lambda\propto T^{\, \beta}$ & \cite{kaz80} \\
$\Lambda\propto e^{-\beta \, a}$ & \cite{raj83} \\
$d\Lambda/dt\propto\Lambda^{\beta}$ & \cite{his86} \\
$\Lambda\propto a^{-2}$ & \cite{lop96,oze86,che90,cal92} \\
$\Lambda\propto a^{-4(1+\epsilon)}$ & \cite{fre87,gas87,sat90,ove93} \\
$\Lambda\propto a^{-m}$ & \cite{ols87,pav91,mai94,sil94,tor96,sil97} \\
$d\Lambda/dt\propto a H^n \Lambda$ & \cite{reu87} \\
$d\Lambda/dt\propto H^3$ & \cite{reu87} \\
$\Lambda\propto C+\beta a^{-m}$ & \cite{sis91,mat95} \\
$\Lambda\propto t^{\, \ell-2}+\beta t^{\, 2(\ell-1)}$ & \cite{kal92} \\
$\Lambda\propto\beta a^{-2}+H^2$ & \cite{car92,arb94} \\
$\Lambda\propto t^{-2}+\beta t^{-2/\ell}$ & \cite{bee93} \\
$\Lambda\propto C+e^{-\beta \, t}$ & \cite{bee93,spi94} \\
$\Lambda\propto C+\beta a^{-2}+H^2$ & \cite{wag93} \\
$\Lambda\propto\beta a^{-m}+H^2$ & \cite{sal93} \\
$\Lambda\propto H^2$ & \cite{lim94a,wet95,arb97} \\
$\Lambda\propto (1+\beta H)(H^2+k/a^2)$ & \cite{lim94b} \\
$\Lambda\propto t^{-1}(\beta +t)^{-1}$ & \cite{kal95} \\
$d\Lambda/dt \propto\beta\Lambda-\Lambda^2$ & \cite{mof96} \\
$\Lambda\propto a^{-3}$ & \cite{hoy97} \\
$\Lambda\propto a^{-2}+\beta a^{-4}$ & \cite{joh97} \\
$\Lambda\propto H^2 + \beta a H (dH/da)$ & \cite{nes97} \\
\end{tabular}
\tablenotetext[1]{$T,a,t,H$ are the temperature, scale factor, time and
                  Hubble parameter respectively;
                  while $\beta,\epsilon,\ell,m$ and $C$ are
                  constants.}
\end{table}

\begin{table}
\caption{B\"orner-Ehlers-type upper limits on matter density $\Omega_0$
         for various values of $m$ and $\gamma$, assuming $z_{\ast}>6$.}
\label{table2}
\begin{tabular}{lcccccccc}
      & $\gamma$: & $2/3$ & $5/6$ & 1 & $7/6$ & $4/3$ & $5/3$ & 2 \\
\tableline
\underline{$m$:} & &       &       &       &       &       &       & \\
$0$   & & $\infty$ & 0.033 & 0.006 & 0.002 & 0.000 & 0.000 & 0.000 \\
$1/4$ & & $\infty$ & 0.16 & 0.10 & 0.079 & 0.067 & 0.053 & 0.044 \\
$1/2$ & & $\infty$ & 0.31 & 0.20 & 0.16 & 0.14 & 0.11 & 0.090 \\
$3/4$ & & $\infty$ & 0.49 & 0.33 & 0.26 & 0.22 & 0.17 & 0.14 \\
1     & & $\infty$ & 0.75 & 0.48 & 0.38 & 0.32 & 0.25 & 0.20 \\
$5/4$ & & $\infty$ & 1.1  & 0.71 & 0.55 & 0.46 & 0.35 & 0.29 \\
$3/2$ & & $\infty$ & 1.9  & 1.1  & 0.87 & 0.71 & 0.54 & 0.44 \\
$7/4$ & & $\infty$ & 4.2  & 2.4  & 1.8  & 1.4  & 1.0  & 0.84 \\
2     & & $\infty$ & $\infty$ & $\infty$ & $\infty$ & $\infty$ & 
          $\infty$ & $\infty$ \\
\end{tabular}
\end{table}


\begin{references}
\bibitem{Absorb} By ``ordinary matter'' we mean matter {\em apart\/} from
                 that corresponding to the $\Lambda$-term.
                 We prefer not to absorb $\Lambda$ into our definitions 
                 of $p$ and $\rho$, as is sometimes done in the literature,
                 so that its effects on the dynamics can be more readily
                 distinguished.
                 The simple equation of state~(\ref{eqState}) is then
                 obeyed by {\em ordinary\/} matter, not the effective 
                 matter obtained by including the $\Lambda$-term.
\bibitem{Exact} Exact solutions can be found, for example, using Jacobian 
                elliptic functions for singular cases with $\gamma=1$ 
                \cite{edw72}.
                Other special cases with $\gamma=1$ can be expressed
                as elliptic integrals \cite{mcv65}.
                Solutions have also been obtained in terms of hypergeometric
                functions for some radiation-like, as well as dust-like 
                cases, by expanding in powers of $\Lambda$ 
                (assumed to be small) \cite{agn70}.
\end{references}
\end{document}